\documentclass[10pt]{iopart}
\usepackage{graphicx,color}
\usepackage[bookmarks]{hyperref}
\usepackage{amssymb}

\newcommand{\be}{\begin{equation}}
\newcommand{\ee}{\end{equation}}

\definecolor{myblue}{rgb}{0,0.482,0.725}

\begin{document}
\review{Dynamical quantum phase transitions: a review}
\author{Markus Heyl}
\address{Max-Planck-Institute for the Physics of Complex Systems, N\"othnitzer Str. 38, 01187 Dresden, Germany}
\ead{heyl@pks.mpg.de}
\date{\today}

\begin{abstract}
Quantum theory provides an extensive framework for the description of the equilibrium properties of quantum matter.  
Yet experiments in quantum simulators have now opened up a route towards generating quantum states beyond this equilibrium paradigm. 
While these states promise to show properties not constrained by equilibrium principles such as the equal a priori probability of the microcanonical ensemble, identifying general properties of nonequilibrium quantum dynamics remains a major challenge especially in view of the lack of conventional concepts such as free energies. 
The theory of dynamical quantum phase transitions attempts to identify such general principles by lifting the concept of phase transitions to coherent quantum real-time evolution.  
This review provides a pedagogical introduction to this field. 
Starting from the general setting of nonequilibrium dynamics in closed quantum many-body systems, we give the definition of dynamical quantum phase transitions as phase transitions in time with physical quantities becoming nonanalytic at critical times.
We summarize the achieved theoretical advances as well as the first experimental observations, and furthermore provide an outlook onto major open questions as well as future directions of research.
\end{abstract}


\submitto{\RPP}
\maketitle
\ioptwocol
\date{\today}

\section{Introduction}
\label{sec:introduction}

Quantum simulators have nowadays achieved experimental access to the real-time dynamics of closed quantum many-body systems due to the impressive progress in controlling matter at the quantum level within the last two decades~\cite{Bloch2008lv,Bloch2012qs,Blatt2012,Georgescu2014}.
Such quantum simulators have been realized on various experimental platforms such as ultra-cold atoms in optical lattices, trapped ions, and more, and have recently studied exotic dynamical phenomena inaccessible with conventional architectures.
This includes the observation of prethermalization~\cite{Gring2012,Neyenhuis2016}, particle-antiparticle production in lattice gauge theories~\cite{Martinez2016}, many-body localization~\cite{Schreiber2015oo,Smith2016,Bordia2016,Choi2016}, or discrete time crystals~\cite{Choi2017,Zhang2017}. 

It is the central property of these nonequilibrium quantum states that they cannot be captured within a thermodynamic description.
This, however, might not be seen as a shortcoming but rather as the defining feature providing the room to realize phenomena not accessible within conventional equilibrium statistical physics.
In turn, conventional strategies, successful for the description of quantum many-body systems in equilibrium, are not applicable.
Concepts such as partition functions or organizing principles such as the minimization of free energies are lacking which provides a significant challenge in approaching a theoretical understanding of nonequilibrium quantum many-body dynamics on general grounds.
This immediately leads to fundamental questions such as whether such systems can nevertheless show universality with macroscopic properties becoming independent of microscopic details?
Is there a dynamical analog of a phase of matter without the existence of a free energy?

In equilibrium systems elementary properties such as universality are intimately connected to the theory of phase transitions~\cite{Sachdev}.
This has motivated various approaches to introduce notions of phase transitions in far-from equilibrium quantum many-body systems~\cite{Eckstein2009wj, Garrahan2010xw, Diehl2010, Schiro2010gj, Sciolla2010jb, Sciolla2011,Ates2012, Sciolla2013, Vosk2014, Wang2016, Maraga2016, Zunkovic2016a,Wang2017, Zhang2017b} addressing qualitative changes in either the long-time dynamics or the asymptotic long-time limit of observables or correlation functions as a function of a microscopic control parameter. 

Remarkably, nonequilibrium phase transitions can also occur on transient time scales with physical quantities becoming nonanalytic as a function of time -- a phenomenon that has been termed Dynamical Quantum Phase Transition (DQPT)~\cite{Heyl2013a}.
Accordingly, DQPTs are driven by progressing time as opposed to conventional phase transitions that are driven by control parameters such as temperature or pressure.
This field has seen substantial progress recently ranging from identifying dynamical order parameters~\cite{Budich2015, Sharma2016, Flaeschner2016, Bhattacharya2017b, Bhattacharya2017,HeylBudich2017}, or  scaling and universality~\cite{Heyl2015dq,Karrasch2017}, to the first experimental observations~\cite{Flaeschner2016,Jurcevic2016}.

It is the aim of this article to introduce pedagocically the concept of DQPTs, to review its key features, to summarize experimental observations, and to identify prospects of the field within a self-contained description.
While the theory of phase transitions is of particular importance for the understanding of the equilibrium properties of matter in nature, it will be one purpose of this review to point out the potential of DQPTs to provide a key principle for the understanding of the \emph{dynamics} in quantum many-body systems.
For a related recent review focusing on DQPTs in exactly solvable model systems see~\cite{Zvyagin2016}.

In the beginning, in section~\ref{sec:dqpts}, a general introduction into the field of DQPTs is given by first introducing the central object termed Loschmidt amplitude and outlining its connection to conventional partition functions which forms the basis for the identification of DQPTs as a nonequilibrium phase transition phenomenon.
In addition, we then present a physical picture of DQPTs as dynamical analogs to equilibrium quantum phase transitions.
Afterwards, in section~\ref{sec:DPTP} we take the particular class of DQPTs occuring in topological systems as an illustrative example providing both a straightforward analytical handle as well as intuitive explanations.
In section~\ref{sec:experiments} the two recent experimental observations of DQPTs in systems of ultra-cold atoms and trapped ions are summarized and put into the theoretical context.
It is the purpose of the subsequent section~\ref{sec:generalPrinciples} to outline how central concepts of equilibrium criticality such as scaling and universality or order parameters extend to DQPTs.
This is followed by a summary of further implications of DQPTs onto other physical quantities, including for example entanglement dynamics, or extensions to a broader range of physical setups such as mixed states, which is presented in section~\ref{sec:furtherApplications}.
The last section~\ref{sec:propects} provides an outlook onto central open questions and potential future research directions in the context of DQPTs.

\section{Dynamical quantum phase transitions}
\label{sec:dqpts}

Within statistical physics the central object for the theoretical description of systems in contact to a heat bath is the (canonical) partition function:
\be
Z = \mathrm{tr} \, e^{-\beta H} = \sum_\nu e^{-\beta E_\nu},
\label{eq:Zdef}
\ee
as a sum of Boltzmann weights over all microstates $\nu$. Here, $H$ denotes the system's Hamiltonian, $\beta$ inverse temperature, and $E_\nu$ the eigenenergies of $H$. The partition function contains the full information about the system's thermodynamic properties because $Z$ is directly related to the free energy $F$ via:
\be
Z = e^{-\beta F} = e^{-\beta N f},
\label{eq:ZF}
\ee
with $f=F/N$ denoting the free energy density and $N$ the number of degrees of freedom.
This equality gives a link between microscopic (relative) probabilities, the Boltzmann weights contained in the partition function, and macroscopic properties, i.e., thermodynamics, through the free energy $F$. 
At a phase transition the thermodynamic potentials such as the free energy $F$ become nonanalytic as a function of the respective control parameter.
When the transition is temperature-driven,  for example, $F$ exhibits a nonanalytic behavior at a critical temperature $T_c$.
This translates directly into nonanalytic structures of the generalized forces and susceptibilities.

\subsection{Closed quantum many-body systems}

As opposed to systems described by equation~(\ref{eq:Zdef}), the focus of this review is on closed quantum many-body systems where the coupling to an environment can be neglected and the dynamics on experimentally relevant time scales can be considered purely unitary and quantum.

Conventional experimental systems such as in the solid state context are naturally embodied with an environment, e.g., phonons serving as a heat bath for the electronic degrees of freedom.
Quantum simulators such as ultracold atoms or trapped ions on the other hand constitute to a high degree of accuracy experimental platforms where the time scales for the system-environment coupling becomes much longer than the internal time scales of the system such that the dynamics can be considered as closed to a very good approximation~\cite{Bloch2008lv,Bloch2012qs,Blatt2012,Georgescu2014}.
While also for solid-state systems pump-probe experiments can induce and detect the dynamics of an approximately closed electronic system~\cite{Fausti2011}, quantum simulators provide a much more tunable and flexible experimental setting.
As such they constitute the platforms most directly connected to the concepts presented in this review article.

\subsection{Nonequilibrium protocol: Quantum quenches}
\label{subsec:quantumQuenches}

In the following we mainly focus on one particular nonequilibrium scenario of a so-called quantum quench.
While the definition of DQPTs is not tied to this specific protocol~\cite{Pollmann2010dv,Heyl2013a,Sharma2015,Divakaran2015,Puskarov2016}, see also section~\ref{subsec:generalProtocols} for a more detailed discussion, quantum quenches are conceptually simplest.

Within a quantum quench the system is initially prepared in the ground state $|\psi_0\rangle$ of an initial Hamiltonian
\be
H_0 =H(\lambda_0) \, ,
\ee
at a value $\lambda_0$ of some tunable parameter $\lambda$ of a more general Hamiltonian $H(\lambda)$.
Then, at a time $t=0$ say, the parameter $\lambda$ is suddenly switched to a new value $\lambda_f$ such that the Hamiltonian has now changed to
\be
H = H(\lambda_f) \, .
\ee
As a consequence the system experiences quantum real-time dynamics that is formally solved by:
\be
|\psi_0(t)\rangle = e^{-iHt} |\psi_0\rangle,
\ee
which will be nontrivial whenever the initial state $|\psi_0\rangle$ is not an eigenstate of the final Hamiltonian $H$. In the remainder, only so-called global quenches will be considered which lead to a macroscopic change in the system's internal energy extensive in system size. This has to be contrasted with local quenches where the energy change is not extensive.

\subsection{Loschmidt amplitudes and Loschmidt echos}
\label{subsec:loschmidt}

The central object within the theory of DQPTs is the Loschmidt amplitude
\be
\mathcal{G}(t) = \langle \psi_0 |\psi_0(t)\rangle = \langle \psi_0 | e^{-iHt} |\psi_0\rangle \, ,
\label{eq:defG}
\ee
which quantifies the deviation of the time-evolved state from the initial condition.
The probability $\mathcal{L}(t)$ associated to the amplitude $\mathcal{G}(t)$:
\be
\mathcal{L}(t) =\big| \mathcal{G}(t) \big|^2\, ,
\label{eq:defL}
\ee
will be termed Loschmidt echo.
Overall, Loschmidt amplitudes and echos appear in various contexts ranging from the theory of quantum chaos~\cite{Peres1984kq,Gorin2006dv}, the Schwinger mechanism of particle production~\cite{Schwinger1951,Martinez2016}, to work distribution functions in the context of nonequilibrium fluctuation theorems~\cite{Talkner2007aw, Silva2008gj, Campisi2011fk,Palmai2015}, among many others.
As such $\mathcal{G}(t)$ and $\mathcal{L}(t)$ represent important quantities in quantum many-body theory.
Within the different anticipated contexts, they appear also under varying terminologies such as return amplitudes, fidelities, or vacuum persistence probabilities.
Because of the formal similarity of Loschmidt amplitudes to partition functions, that will be discussed in detail below in section~\ref{subsec:fisherZeros}, they experience a particular functional dependence on  the number of degrees of freedom $N$  in the limit of large $N$ for the considered global quantum quenches~\cite{Heyl2013a,Gambassi2012a}:
\be
\mathcal{G}(t) = e^{-N g(t)}\, ,
\ee
with $g(t)$ the associated rate function.
Alternatively, the above scaling might also be rephrased in the form that
\be
g(t) =-\lim_{N\to\infty} \frac{1}{N} \log\big[ \mathcal{G}(t) \big]\, ,
\ee
has a well-defined thermodynamic limit.
Analogously, for the Loschmidt echo, let us introduce the rate function $\lambda(t)=-\lim_{N\to\infty} N^{-1} \log \mathcal{L}(t) $ such that :
\be
\mathcal{L}(t) = e^{-N\lambda(t)} \, ,
\label{eq:def_LE_rateFunction}
\ee
with $\lambda(t) = 2\mathrm{Re}[g(t)]$.
This typical large-deviation scaling~\cite{Touchette} of $\mathcal{G}(t)$ and $\mathcal{L}(t)$ with exponential dependence on system size $N$ can change for critical initial states when the quantum quench induces superextensive energy fluctuations in the system~\cite{Heyl2017a}.
But for the sake of the review we restrict to cases where the conventional large-deviation scaling emerges.

Since in our quantum quench protocol the initial condition is fixed to be the ground state of the initial Hamiltonian, see section~\ref{subsec:quantumQuenches}, the Loschmidt echo requires a generalization whenever the ground state is degenerate, e.g., in case of symmetry-broken phases.
While such a generalization is not unique, the following one has turned out to be very fruitful~\cite{Jurcevic2016,Heyl2014,Zunkovic2016,Weidinger2017}.
Interpreting the Loschmidt echo as the probability to return to the ground-state manifold, the natural extension for the case of a discrete symmetry is given by:
\be
	P(t) = \sum_{\alpha=1}^\nu P_\alpha(t),\quad P_\alpha(t)= \big| \langle \psi_\alpha | \psi(t)\rangle \big|^2 \, ,
	\label{eq:groundStateProb}
\ee
with $|\psi_\alpha \rangle$ the set of $\nu$ degenerate ground states of the initial Hamiltonian.
For $\nu=1$ this directly reproduces the definition of the Loschmidt echo in equation~(\ref{eq:defL}). 
When the system exhibits a continuous symmetry, a generalization following the same spirit as in equation~(\ref{eq:groundStateProb}) has been recently proposed replacing the sum by an integral over the manifold of degenerate ground states~\cite{Weidinger2017}.
Notice that DQPTs cannot only happen in the full probability $P(t)$, but also in the individual $P_\alpha(t)$~\cite{Halimeh2016, Homrighausen2017, Halimeh2017pa}

\subsection{Definition of dynamical quantum phase transitions}
\label{subsec:defDQPT}

Now let us come to the actual definition of a DQPT.
As anticipated before, phase transitions in equilibrium are accompanied by a nonanalytic structure of the free energy upon varying the control parameter.
In close analogy, in the following we identify a DQPT as a nonanalytic behavior of the Loschmidt amplitude as a function of time.
In this sense, DQPTs are dynamical in nature and can thus be thought of as \textit{phase transitions in time}.
%
%
%

\begin{figure}
\centering
\includegraphics[width=0.95\columnwidth]{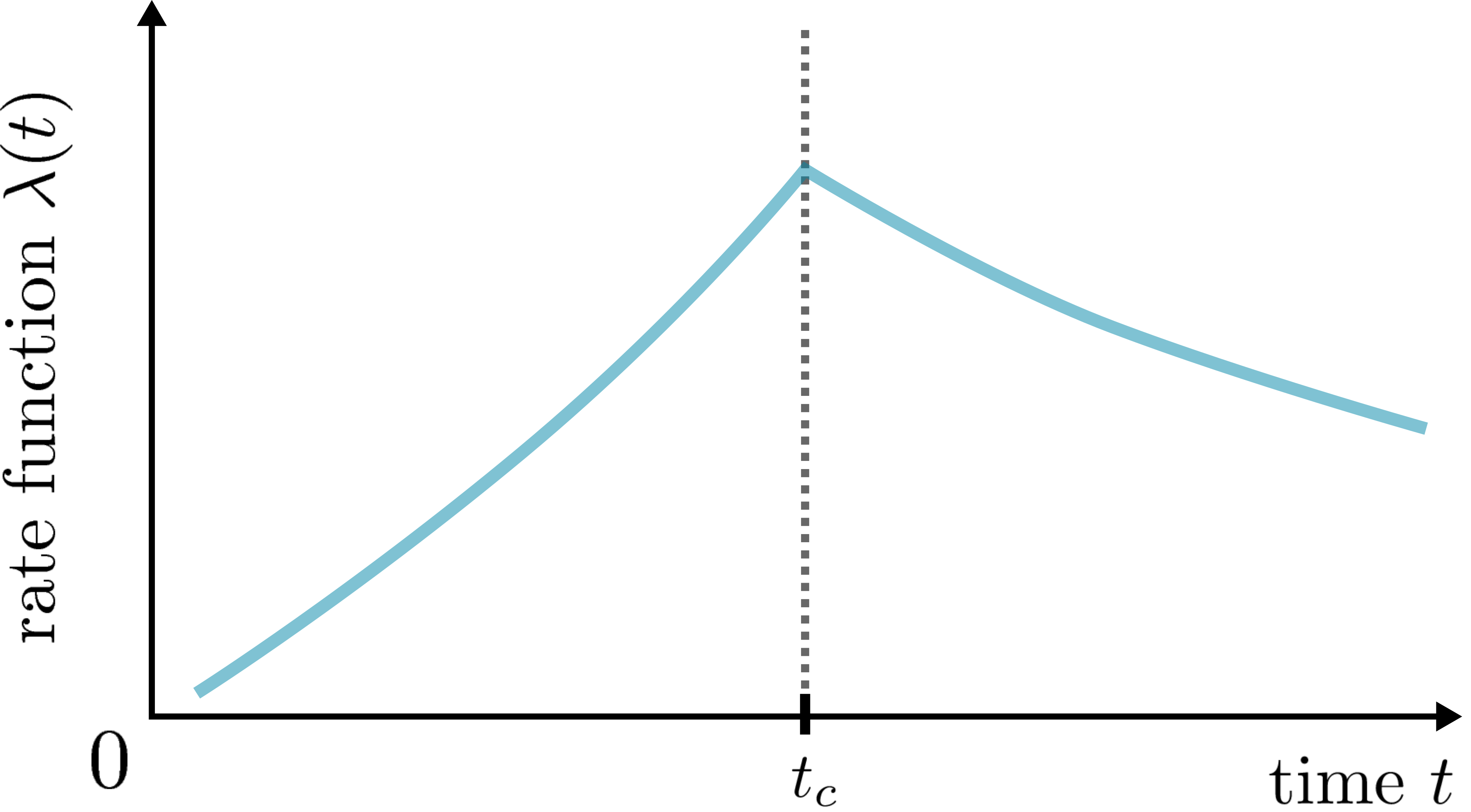}
\caption[DQPT schematic illustration]{Schematic illustration of a dynamical quantum phase transition. At a critical time $t=t_c$ the rate function $\lambda(t)$ of the Loschmidt echo exhibits a nonanalytic kink. While these kinks occur in many exactly-solvable one-dimensional models, in particular in higher dimensions the nonanalytic structure can be different.}
\label{fig:dqptSchematic}
\end{figure}

In figure~\ref{fig:dqptSchematic} an illustration of a prototypical example of a DQPT is shown as it appears in various systems.
In this schematic sketch, DQPTs are associated with a kink in the rate function $\lambda(t)$ of the Loschmidt echo yielding the following functional behavior in the vicinity of the critical time $t_c$:
\be
        \lambda(t) \sim \left|\frac{t-t_c}{t_c} \right| \, .
\ee
Such kinks appear in various one-dimensional (1D) systems. However, especially in two dimensions (2D) different nonanalytic structures have been found.
For quantum quenches in Chern insulators, for example, power-law nonanalyticities can emerge~\cite{Vajna2015} and in 2D Ising models logarithmic singularities~\cite{Heyl2015dq}.
While nonanalytic real-time behavior of Loschmidt amplitudes has been recognized already by Pollmann et al.~\cite{Pollmann2010dv}, the interpretation as a dynamical critical phenomena has been first pointed out by~\cite{Heyl2013a}.
Under which circumstances can DQPTs appear?
Overall, it has been observed in most of the reported cases that DQPTs occur whenever a parameter of the Hamiltonian is quenched across an underlying equilibrium transition.
However, notable exceptions exist~\cite{Vajna2014, Canovi2014fo, Andraschko2014, Schmitt2015, Zunkovic2016, Halimeh2016} suggesting that DQPTs are not in a one-to-one correspondence to conventional phase transitions, for a more detailed discussion see section~\ref{subsec:eqtransitions}.
Thus, DQPTs should, generally speaking, be rather seen as a critical phenomenon distinct from the equilibrium case.
As anticipated already in the introduction, also other notions of dynamical phase transitions have been reported in the literature~\cite{Schutzhold2006,Eckstein2009wj, Garrahan2010xw, Diehl2010, Schiro2010gj, Sciolla2010jb, Sciolla2011, Ates2012, Sciolla2013,  Vosk2014,Wang2016, Maraga2016, Zunkovic2016a,Wang2017, Zhang2017b}, which in some cases can be linked to DQPTs~\cite{Zunkovic2016,Wang2017,Weidinger2017}.
We discuss these and other notions as well as some of the connections to DQPTs in more detail in section~\ref{subsec:otherdpts}.
In this review, we resort to the definition in terms of Loschmidt amplitudes given above.

\subsection{Complex partition functions and Fisher zeros}
\label{subsec:fisherZeros}

Having given the definition of DQPTs it is the aim of the following section to address the elementary question of why it is possible that Loschmidt amplitudes can become nonanalytic as a function of time.
Associated with that: is this a generic feature or does this require fine-tuning?
To see that DQPTs can occur generically without fine-tuning, it is most straightforward to resort to an important concept of equilibrium phase transitions: complex partition function zeros also known as Fisher~\cite{Fisher1967da} or Lee-Yang zeros~\cite{YangLee1,YangLee2}.
While these complex zeros have been originally introduced as a purely theoretical concept, it is worthwhile to emphasize that recently it became possible to measure them experimentally~\cite{Peng2015eo, Brandner2017}.

In order to apply this concept to the problem at hand here, let us first point out that there is a specific class of equilibrium partition functions that shares an immediate formal similarity to Loschmidt amplitudes.
Consider an equilibrium system with boundary conditions imposed on two ends a distance $R$ apart from each other.
Then, importantly for the present aim, the respective so-called boundary partition function $Z_B$ can be represented in the following form:~\cite{LeClair1995ho}
\be
Z_B = \langle \psi_1 | e^{-RH} | \psi_2 \rangle,
\ee
with the states $|\psi_1\rangle$ and $|\psi_2\rangle$ encoding the boundary conditions and $H$ denoting the bulk Hamiltonian.
Thus, formally, Loschmidt amplitudes can be identified with boundary partition functions at a complexified parameter $R =it$.
Accordingly, the initial state $|\psi_0\rangle$ plays the role of a boundary condition in time instead of space.
This identification is not only useful for the subsequent general discussion, it has also been utilized as a computational tool to construct a complex transfer matrix representation of Loschmidt amplitudes~\cite{Andraschko2014} which can be efficiently computed using DMRG methods in low dimensions~\cite{Andraschko2014,James2015}.

Although, in general, a partition function with complex parameters does not describe a physical system, the complexification of otherwise real parameters in partition functions has been a central concept for the theory of phase transitions~\cite{Fisher1967da}, which we can now also apply directly to Loschmidt amplitudes.
\begin{figure}
\centering
\includegraphics[width=0.9\columnwidth]{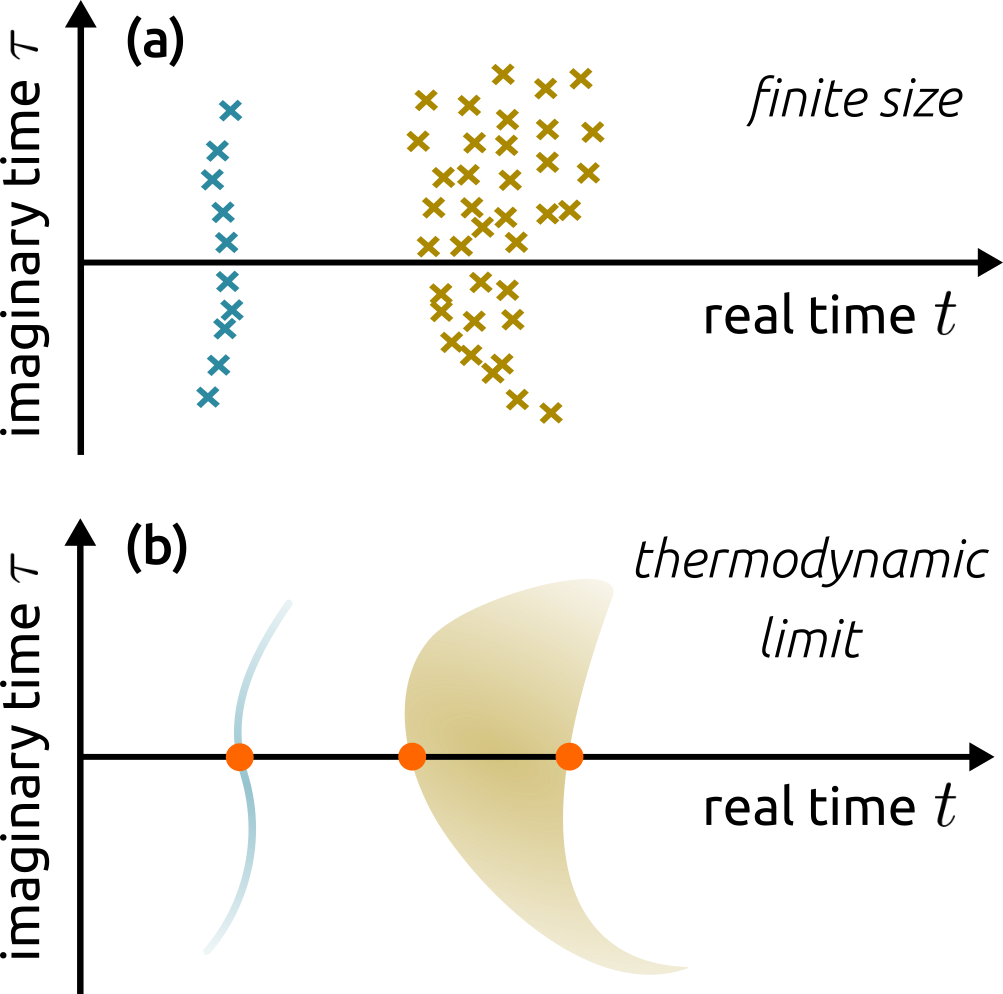}
\caption[Fisher zeros]{Schematic illustration of Fisher zeros in the complex time plane. {\bf (a)} For systems of finite size, the zeros appear as points in the complex parameter plane. {\bf (b)} Upon increasing system size, Fisher zeros start to accumulate and form structures which can generically be of two different kinds. First, and this happens often in low-dimensional systems, Fisher zeros can coalesce onto lines. Second, Fisher zeros can accumulate to form areas. DQPTs occur whenever a such a line or boundary of an area hits the real-time axis indicated by the dots along the real-time axis.}
\label{fig:fisherZeros}
\end{figure}
Consider the Loschmidt amplitude with time $t \mapsto z =t +i\tau \in  \mathbb{C}$ extended into the complex plane:
\be
\mathcal{G}(z) = \langle \psi_0 | e^{-iHz} | \psi_0 \rangle, \quad z\in\mathbb{C}.
\ee
For a finite-sized system partition functions or equivalently Loschmidt amplitudes are analytic functions.
One way to see this for systems composed out of spins or fermions on a lattice, is to insert an eigenbasis $|E_\nu\rangle$ of the Hamiltonian $H$ with corresponding energies $E_\nu$ such that:
\be
	\mathcal{G}(z) = \sum_\nu \big| \langle E_\nu | \psi_0\rangle \big|^2 e^{-i E_\nu z}
\ee
Because for fermionic or spin systems the eigenbasis is finite for $N<\infty$ we have for $\mathcal{G}(z)$ a finite sum of analytic functions which results itself in an analytic function.
As a consequence, the Weierstrass factorization theorem~\cite{Conway} applies which allows us to represent $\mathcal{G}(z)$ via:
\be
\mathcal{G}(z) = e^{\mu(z)} \prod_{j} \left[ z_j-z \right],
\ee
where the $z_j$ denote the zeros of $\mathcal{G}(z)$ in the complex plane. 
While $\mu(z)$ is, by the theorem, always an analytic function, all the nonanalytic properties of $\mathcal{G}(z)$ in the thermodynamic limit are contained in the structure of the zeros $z_j$ in the complex plane.
Disregarding for the moment the smooth function $\mu(z)$, the singular contribution $g_s(t)$ to the rate function of the Loschmidt amplitude is:
\be
        g_s(z) = -\frac{1}{N} \sum_j \log \big[ z_j-z \big].
\ee
As for conventional equilibrium partition functions, the zeros $z_j$ represent isolated points in the complex plane for $N<\infty$, for an illustration see figure~\ref{fig:fisherZeros}.
Upon increasing system size $N$, however, the zeros accumulate on lines or areas depending on the details of the system~\cite{Saarloos1984}.
Why is the concept of Fisher zeros now important for DQPTs?
This is because nonanalyticities and thus phase transitions occur whenever a line or a boundary of an area of Fisher zeros is crossed in course of time evolution.
On a general level one can see this by noticing an interesting connection between electrostatics and the real part of $g_s(z)$~\cite{Schmitt2015}, i.e., the singular part of the rate function $\lambda_s(z) = 2\mathrm{Re}[g_s(z)]$ of the Loschmidt echo, see equation~(\ref{eq:def_LE_rateFunction}).
Defining an effective density $\rho(z)$ of Fisher zeros via
\be
	\rho(z) = \frac{2}{N} \sum_j \delta\Big( z-z_j \Big)
\ee
it is possible to represent $\lambda_s(t)$ as:
\be
	\lambda_s(z) = -\int_{\mathbb{C}} d\overline{z} \, \rho(\overline{z}) \log \left| \overline{z} - z \right|.
\ee
Because $\log|z|$ is the Green's function of the two-dimensional Laplacian $\Delta = \partial_t^2+\partial_\tau^2$ with $z=t+i\tau$, we can interpret $\lambda_s(z)$ as an electrostatic potential that is generated by an effective charge density $\rho(z)$.
Nonanalyticities in electrostatic potentials and therefore equivalently in $\lambda_s(z)$ occur at nonanalytic structures of the respective charge density $\rho(z)$.
One can distinguish two different cases, see figure~\ref{fig:fisherZeros}.
First, when the Fisher zeros fall onto a line, the nonanalytic behavior of $\lambda_s(t)$ is determined solely by the line density of defects~\cite{Fisher1967da}.
Second, when the Fisher zeros form an area, the nonanalytic structure becomes equivalent to that of an electrostatic potential at surfaces between two regions of different charge densities~\cite{Schmitt2015}.

In the context of complex partition functions it might be interesting to study also the relation to another notion of dynamical phase transitions~\cite{Rotter2010,Rotter2015}, which occurs for so-called exceptional points of non-Hermitian Hamiltonians at complex parameters, where connections to DQPTs might appear naturally.
Along these lines, a further interesting aspect of partition functions at complex temperatures is that it has been shown that they can be represented in a form of the Loschmidt amplitude $\mathcal{G}(t)$ using so-called canonical states~\cite{borrmann2000classification}, which up to now has not been explored.

While the robustness of DQPTs will be summarized in more detail in section~\ref{sec:robustness}, the Fisher zero considerations of the present section provide the opportunity for a preliminary short discussion.
From equilibrium it is known that phase transitions are robust against symmetry-preserving perturbations that are weak in the renormalization group sense.
From the perspective of partition function zeros this means that the respective structures, lines or areas, are also robust.
Specifically, symmetry-preserving perturbations might deform these structures, but provided they are sufficiently weak, these deformations don't lead to a melting of the lines of areas of zeros.
This robustness has indeed been observed for DQPTs both analytically~\cite{Kriel2014} as well as numerically~\cite{Karrasch2013,Kriel2014,Heyl2015dq, Sharma2015}.
If, however, the perturbation breaks a symmetry of the system, the stability is not guaranteed as has also been seen numerically for particular cases~\cite{Pollmann2010dv,Obuchi2017}.
In this sense, the robustness of DQPTs seems to follow similar principles as at conventional phase transitions for the examples in the literature.

\subsection{Analogy to equilibrium quantum phase transitions}
\label{subsec:QPT}

\begin{figure}
\centering
\includegraphics[width=1\columnwidth]{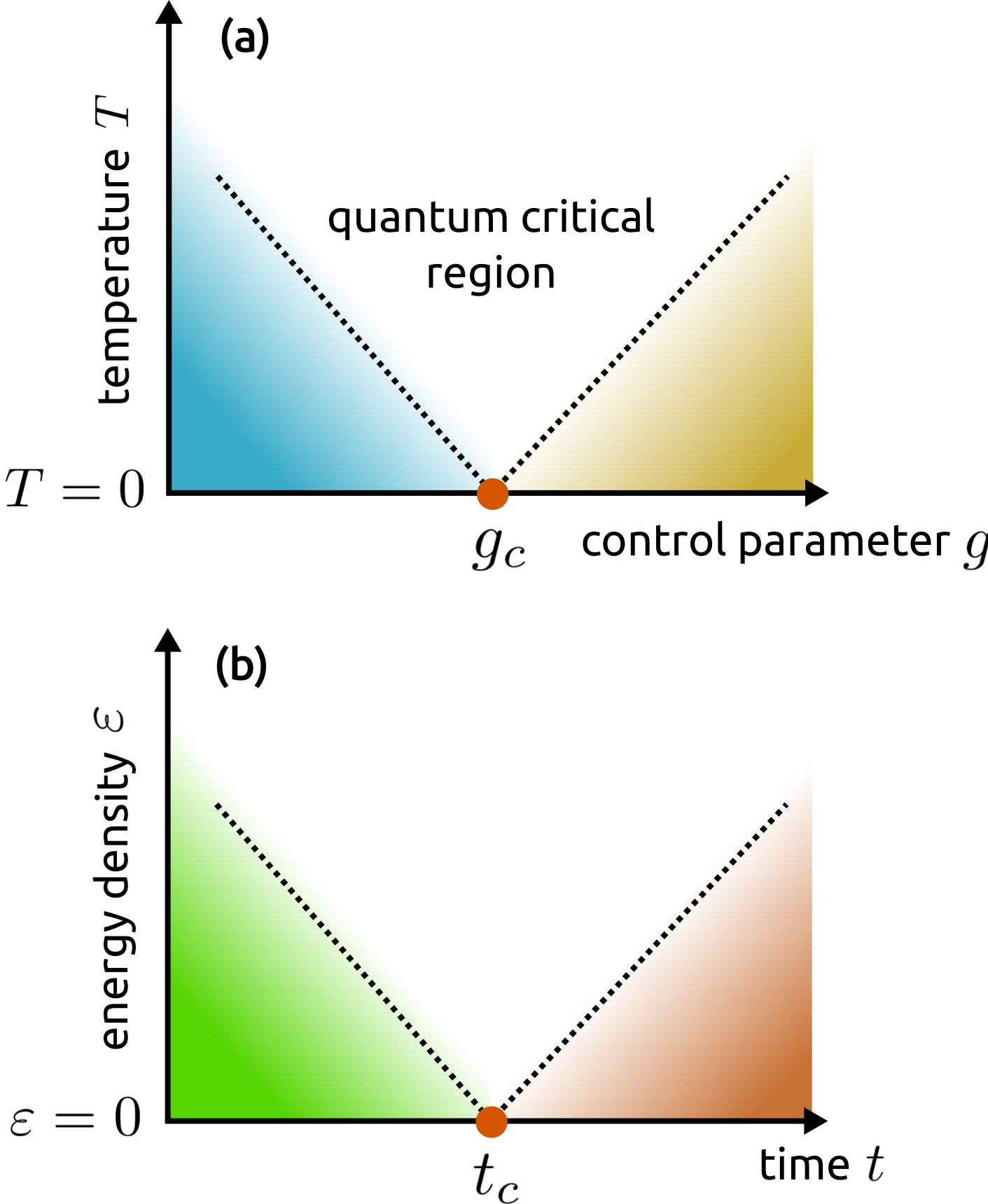}
\caption[Analogy to quantum phase transitions]{Schematic illustration of the analogy between dynamical quantum phase transitions and conventional quantum phase transitions. {\bf (a)} A quantum phase transition occurs at zero temperature $T=0$ at a critical value $g_c$ of a quantum control parameter $g$. While at nonzero temperatures $T>0$ the quantum phase transition disappears, there emerge two crossover lines instead enclosing the quantum critical region whose properties are controlled by the underlying critical point. {\bf{(b)}} A dynamical quantum phase transition is driven by time $t$ with Loschmidt amplitudes becoming nonanalytic  at a critical time $t=t_c$. The Loschmidt amplitude probes the ground-state manifold of the initial Hamiltonian (energy density $\varepsilon=0$). While the nonanalytic behavior can disappear for excited energy densities $\varepsilon>0$, where local observables acquire their dominant contribution, there can still be an extended region (white space) controlled by the underlying dynamical critical point.}
\label{fig:qpt}
\end{figure}

It is the goal of the following discussion to give a physical interpretation of DQPTs in terms of a dynamical analog to conventional quantum phase transitions (QPT)~\cite{Heyl2014},  see figure~\ref{fig:qpt} for an illustration.
This interpretation aims at providing a general argument of how DQPTs can control the dynamics of other observables.
Since the Loschmidt amplitude $\mathcal{G}(t)$ is a projection of the full time-evolved many-body wave function $|\psi_0(t)\rangle$ onto one single state in Hilbert space (the initial state) and therefore only retrieves partial information of $|\psi_0(t)\rangle$, one might wonder to which extent the overlap $\mathcal{G}(t)$ can be important for the understand of the dynamics of the full wave function $|\psi_0(t)\rangle$.
Overall, this amounts to the question whether this single overlap represents a singular point or whether also overlaps with other states in Hilbert space show similar properties such that in this sense the properties of the Loschmidt amplitude can spread out to larger portions of Hilbert space.
QPTs represent an important example where the singular behavior in a single many-body state, the ground state, can influence an extended set of states within the quantum critical region at nonzero temperature $T>0$.
In the following, we argue that an analogous picture can emerge also for DQPTs, and be made concrete and even measurable, see section~\ref{subsec:expIons}.

As already anticipated before, the Loschmidt amplitude $\mathcal{G}(t)=\langle \psi_0 | \psi_0(t) \rangle$ is a projection of the time-evolved state $|\psi_0(t)\rangle$ back onto the initial state $|\psi_0\rangle$, which is always chosen as the ground state of the initial Hamiltonian.
From this perspective, Loschmidt amplitudes probe the asymptotic low-energy properties of $|\psi_0(t)\rangle$ when measuring energies with the \emph{initial} and not the final Hamiltonian~\cite{Heyl2014}.
In this sense, the nonanalyticities associated with DQPTs are a ground-state manifold property in close analogy to conventional QPTs occurring at zero temperature $T=0$~\cite{Heyl2014}.
This interpretation naturally leads to the general picture in figure~\ref{fig:qpt}.
Instead of representing the phase diagram in the temperature-control parameter plane in the case of a QPT, at a DQPT one might think of an energy density-time plane with energy density $\varepsilon$ replacing temperature $T$ and time $t$ the control parameter $g$.
In that picture the DQPT occurs along the line of vanishing energy density $\varepsilon=0$ (upon choosing the zero of energy accordingly) at a critical time $t_c$.

Due to the quantum quench, however, energy is pumped into the system and the dominant contribution to local observables or correlation functions is originating from a narrow shell in the vicinity of the mean energy density $\varepsilon_\mathrm{av}(t) = N^{-1} \langle H_0 (t) \rangle$~\cite{Heyl2014} beyond the ground-state manifold.
The central question in the end becomes whether there exists a dynamical analog to a quantum critical region controlled by the DQPT $\varepsilon=0$ and whether $\varepsilon_\mathrm{av}(t)$ crosses that region or not.
For certain examples, this ascribed analogy to QPT can be made concrete~\cite{Heyl2014} and even measured experimentally~\cite{Jurcevic2016}, as will be discussed in more detail in section~\ref{subsec:expIons}.
Whether, however, such a dynamical analog of a quantum critical region exists for any DQPT is not known and has to be checked on a case-by-case basis.

The interpretation of DQPTs as a dynamical analog to QPT might also be extended further.
In the equilibrium case, QPTs cannot be observed directly in experiments because of the third law of thermodynamics.
DQPTs also exhibit an analog to the third law in the sense that it is not possible to experimentally observe them without further theoretical input, as it was used in the recent trapped ion experiment~\cite{Jurcevic2016}.
This is because of the exponential suppression of, e.g., the Loschmidt echo $\mathcal{L}(t)$, with the number of degrees of freedom $N$:
\be
	\mathcal{L}(t) = e^{-N\lambda(t)}.
\ee
Importantly, $\lambda(t)$ is independent of $N$ in the thermodynamic limit implying that it becomes exponentially challenging and therefore asymptotically impossible to measure $\mathcal{L}(t)$ experimentally.
Since DQPTs only occur for $N\to\infty$, observing them in an experiment becomes exponentially hard.
\subsection{Relation to equilibrium phase transitions}
\label{subsec:eqtransitions}
It has been observed in many cases that DQPTs share a close connection to underlying equilibrium QPTs.
It therefore appears as a central question how these two phase transition phenomena are related to each other.
For topological systems of noninteracting fermions the connection is by now particularly clear for two-band models~\cite{Vajna2015, Huang2016}, as will be discussed in more detail also in section~\ref{subsec:protectedaccidental}.
Whenever a topological phase transition is crossed by a quantum quench in 1D, a DQPT necessarily has to emerge.
In 2D the situation is a bit more subtle and requires the absolute value of the Chern number of the underlying equilibrium ground states to change.
While these DQPTs are topologically protected~\cite{Vajna2015}, also so-called accidental DQPTs can occur without crossing a QPT.
A similar phenomenology has been observed for the $XY$ chain in a transverse field~\cite{Vajna2014} that is also mappable to a system of noninteracting fermions in 1D.
In addition, however, it was found that for this model it can also occur that no DQPT arises even though the quench crossed an underlying equilibrium QPT.
This particular property can be traced back to a kinetic constraint, as also observed, for example, for the ferromagnetic $XXZ$ chain~\cite{Andraschko2014}.
This kinetic constraint is a $U(1)$ symmetry due to particle or magnetization conservation which does not allow to dynamically enter the particle number or magnetization sectors the system adopts in the equilibrium case.
Without a coupling to a grand-canonical bath with particle-number exchange, the system is there trapped in a fixed sector which is lifting in general the connection between the dynamical and equilibrium phase transitions.
All these examples are related to systems which don't exhibit nonzero-temperature phase transitions such that order only exists in the ground states of the respective models.
How order in excited states, such as relevant for so-called excited state phase transitions~\cite{Cejnar2007, Caprio2008,ExcState1,Santos2015,Santos2016, Stransky2016, Sindelka2017},  affects DQPTs is a much more intricate question and the situation is much less clear at this point mainly because there exist only a few studied models in the literature with such properties~\cite{Heyl2015dq, Zunkovic2016, Halimeh2016, Halimeh2017pa, Homrighausen2017, Canovi2014fo}.
For quenches in the 2D transverse-field Ising model it has been found that DQPTs emerge with the same nonanalytic structures as the free energy at the equilibrium nonzero-temperature critical point of the classical 2D Ising model~\cite{Heyl2015dq}.
This suggests again a close connection of DQPTs to equilibrium phase transitions.
In long-range interacting Ising models on the other hand it has been found that the various observed DQPTs are not related to an underlying equilibrium phase transition~\cite{Zunkovic2016, Halimeh2016, Halimeh2017pa, Homrighausen2017}, i.e., neither the quantum nor the nonzero-temperature one.
While a connection between DQPTs and another nonequilibrium phase transition in the long-time steady state after the quench has been observed~\cite{Zunkovic2016, Halimeh2016}, evidence also for DQPTs in a regime of a weak quench has been found~\cite{Halimeh2016,Halimeh2017pa,Homrighausen2017}, which have been termed 'anomalous' accordingly~\cite{Halimeh2016}.
These anomalous DQPTs connect also to an observation in other models where it has been shown that DQPTs can occur even without crossing an underlying equilibrium transition~\cite{Andraschko2014,Vajna2014,Schmitt2015}.

Summarizing, in many cases there appears a strong connection between DQPTs and underlying equilibrium phase transitions, especially in low dimensions.
However, the results in the literature also show that a substantial number of counterexamples  exist, which suggest that DQPTs constitute a genuine nonequilibrium phenomenon.

\section{Dynamical topological quantum phase transitions}
\label{sec:DPTP}

The theory of DQPTs in noninteracting topological systems, termed also dynamical \emph{topological} quantum phase transitions, has reached a rather extensive understanding in recent years~\cite{Hickey2014,Vajna2015,Schmitt2015, Budich2015, Huang2016, Flaeschner2016, Bhattacharya2017, Bhattacharya2017c, Bhattacharya2017b, HeylBudich2017}.
Interestingly, DQPTs in these models are strongly connected to the underlying equilibrium topological properties: In 2D, for example, DQPTs always appear for quenches between two Hamiltonians with different absolute value of the Chern number~\cite{Vajna2015,Huang2016}.
Moreover, DQPTs in such topological systems can be characterized by dynamical order parameters~\cite{Budich2015, Flaeschner2016, Bhattacharya2017, Bhattacharya2017b} which are capable to distinguish the two 'dynamical phases' separated by a DQPT.
As will be discussed in section~\ref{subsec:expAtoms}, for a 2D system such an order parameter has in the meantime been measured in an ultra-cold atom experiment.
Overall, quantum quenches in these topological systems provide an instructive example offering both intuitive explanations and straightforward mathematical understanding of the nature and occurences of DQPTs.
Many of the discussed formal properties, as long as they are not of topological origin, also directly extend to other fermionic systems  or spin models that are mappable to fermionic ones~\cite{Zvyagin2015,Zvyagin2016,Zvyagin2017,Hickey2014b}, which are alternatively summarized in the anticipated recent review~\cite{Zvyagin2016}.

\subsection{Two-band models}
\label{subsec:twobandmodels}

For the sake of simplicity we study DQPTs for two-band topological systems.
For extensions to multiple bands we refer to~\cite{Huang2016}.
Consider noninteracting fermions exhibiting translational invariance and particle-hole symmetry.
Such systems exhibit a compact representation of the Hamiltonian
\be
	H = \sum_{k \in \mathrm{BZ}} H_k, \quad H_k = \mathbf{c}_k^\dag h_k \mathbf{c}_k \, ,
\ee
with the momentum summation extending over the Brioullin zone (BZ).
Here, $\mathbf{c}_k$ denotes a spinor which has different representations depending on the microscopic details of the studied model system.
This spinor can be of the form $\mathbf{c}_k = (c_{k A},c_{kB})$ for insulators with $A$ and $B$ referring, for example, to two sublattices, or the spinor can acquire the form  $\mathbf{c}_k = (c_k,c_{-k}^\dag)$ for superconductors. 
The properties of the particular model are fully specified  by the Hermitian $2\times 2$ matrices $h_k$ which can be represented in terms of Pauli matrices $\tau_\alpha, \alpha=x,y,z$, due to particle-hole symmetry in the following form:
\be
	h_k = \mathbf{d}_k \mathbf{\tau} = \sum_{\alpha=x,y,z} d_k^\alpha \tau_\alpha \, .
	\label{eq:defhk}
\ee
For a 1D Kitaev chain, for example, one has that $\mathbf{d}_k = [0,\Delta \sin(k),\mu-J\cos(k)]$ with $J$ denoting the hopping strength, $\Delta$ the pairing amplitude, and $\mu$ the chemical potential.
The Hamiltonian can be diagonalized for each momentum sector $k$ separately yielding the two Bloch states $|\psi_{k+}\rangle$ and $|\psi_{k-}\rangle$ with energies $+\varepsilon_k$ and $-\varepsilon_k$, respectively, with $\varepsilon_k = |\mathbf{d}_k| $.
Because the different momentum sectors are decoupled, ground states (as well as other eigenstates) exhibit a factorization property:
\be
	|\psi\rangle = \prod_k |\psi_{k-}\rangle \, .
\ee
Moreover, any nonequilibrium protocol which induces a time-dependence in $\mathbf{d}_k(t)$ without coupling of the momentum sectors preserves this property yielding
\be
	|\psi(t) \rangle = \prod_k |\psi_{k-}(t) \rangle,
	\label{eq:factorizationState}
\ee
which has the advantage that the dynamics can be studied for each momentum $k$ separately.

For the considered case of a quantum quench, the problem is fully specified by the vectors $\mathbf{d}_k^{i/f}$ for the initial/final Hamiltonian, respectively.
Accordingly, let us denote the corresponding Bloch states by $|\psi_{k\pm}^{i/f}\rangle$ and the energies via $\varepsilon_k^{i/f} = |\mathbf{d}_k^{i/f}|$.

\subsection{Loschmidt amplitude}
\label{subsec:dtptle}

Based on the considerations of the previous section, it is straightforward to study DQPTs in the Loschmidt amplitude.
Due to the factorizing property of the quantum many-body state in equation~(\ref{eq:factorizationState}), $\mathcal{G}(t)$ also factorizes
\be
	\mathcal{G}(t) = \prod_k \mathcal{G}_k(t), \quad \mathcal{G}_k(t) = \langle \psi_{k-}^i | \psi_{k-}^i(t)\rangle \,.
	\label{eq:factorizationG}
\ee
In order to evaluate this expression it is suitable to introduce the occupations
\be
	n_k^f = | \langle \psi_{k+}^f | \psi_{k-}^i \rangle|^2 \, ,
\ee
of the upper Bloch band of the final Hamiltonian, which is a constant of motion during the nonequilibrium quantum evolution.
Expanding $|\psi_{k-}^i\rangle$ in the Bloch states of the final Hamiltonian $| \psi_{k\pm}^f \rangle$ one obtains
\be
	\mathcal{G}_k(t) = n_k^f e^{i\varepsilon_k^f t} + (1-n_k^f) e^{-i\varepsilon_k^f t}.
	\label{eq:Gk}
\ee
As discussed in section~\ref{subsec:fisherZeros}, nonanalytic behavior in Loschmidt amplitudes and thus DQPTs is associated with zeros  of $\mathcal{G}(t)$.
Because of equation~(\ref{eq:factorizationG}) a zero in $\mathcal{G}(t)$ is equivalent to finding at least one critical momentum $k^\ast$ and one critical time $t_c$ where $\mathcal{G}_{k^\ast}(t_c)=0$.
According to equation~\ref{eq:Gk}, such a zero can occur whenever there is a mode $k^\ast$ with~\cite{Heyl2013a}
\be
      n_{k^\ast}^f = \frac{1}{2} \, .
      \label{eq:conditionCriticalMomentum}
\ee
Then, there is a series of critical times
\be
	t_c^n = (2n+1) \frac{\pi}{2\varepsilon_{k^\ast}^f}\, ,
	\label{eq:criticalTimes}
\ee
for which $\mathcal{G}_{k^\ast}(t_c^n)=0$.
Using this insight, the question remains under which conditions $n_{k^\ast}^f = 1/2$ is possible.
Formally, it means that the two-level system at $k^\ast$ is maximally mixed, i.e., equivalent to an infinite temperature state.
In general, the occurrence of a critical momentum $k^\ast$ depends on the details of the studied problem. 
However, in many systems the existence of a $k^\ast$ is ensured whenever the system is quenched across an underlying equilibrium quantum critical point as discussed in the following.

\subsection{Landau-Zener problem}
\label{subsec:landauzener}

Why the crossing of an underlying equilibrium quantum phase transition can lead to the appearance of DQPTs can be seen most directly by invoking general Landau-Zener arguments~\cite{Kolodrubetz2012}.
For that purpose, let us first consider a more general scenario of a parameter ramp $\lambda(t) = vt + \lambda_0$ for our general Hamiltonian $H(\lambda)$ which will be used afterwards to argue about the quantum quench case.
Here, $\lambda_0$ and $v$ are chosen such that $\lambda(t)$ interpolates between the initial ($\lambda_0$) and final ($\lambda$) values of the parameter from $t=0$ to $t=\tau$ where $\tau$ is the ramp time.
Let us focus on the situation where the ramp crosses an underlying equilibrium quantum critical point with a gap closing.

Since all momenta $k$ are decoupled from each other we can study the parameter ramp problem for each $k$ separately.
For each $k$ we are dealing with a two-level system such that we can define a momentum-dependent gap $\Delta_k(\lambda)$.
Starting with a slow ramp, the threshold for the breakdown of adiabaticity and therefore exciting to the upper Bloch band is~\cite{Polkovnikov2011kx}
\be
	\frac{d}{dt} \Delta_k[\lambda(t)] = \Delta^2_k[\lambda(t)] \, .
\ee
For the momentum $k_0$ exhibiting the gap closing  this threshold is crossed with certainty and the excitation probability into the upper of the two levels approaches unity, i.e., $n_{k_0}^f \to 1$ implying almost complete occupation inversion~\cite{Kolodrubetz2012}.
On the other hand, in most cases there are modes $k$ that are only weakly excited with $n_k^f \ll 1$.
By using continuity there has then to be at least one momentum $k^\ast$ for which $n_{k^\ast}^f=1/2$ accordingly, which is the required condition for the presence of DQPTs.
Importantly, the general picture for the ramp naturally extends to quantum quenches by decreasing the ramp time $\tau$.
It appears that making the ramp faster and therefore exciting the system stronger is typically not changing the final occupation of the $k_0$ mode which has already reached its maximum value.
What can happen is a shift of the critical mode $k^\ast$ which, however, only modifies the time scales of appearance of DQPTs, see equation~(\ref{eq:criticalTimes}), but not the principle occurrence.

Consequently, a gap closing, i.e., crossing an underlying quantum phase transition, is often a sufficient condition for obtaining a DQPT.
This argument might not be applicable to cases where all $k$ modes are strongly excited with $n_k^f>1/2$, which occurs, for example, when inverting the complete band structure.
In most generic situations, this scenario, however, does not occur and occupation inversion happens only for a subset of modes.

\subsection{Relative Bloch sphere}
\label{subsec:relativeBlochSphere}

\begin{figure}
\centering
\includegraphics[width=0.7\columnwidth]{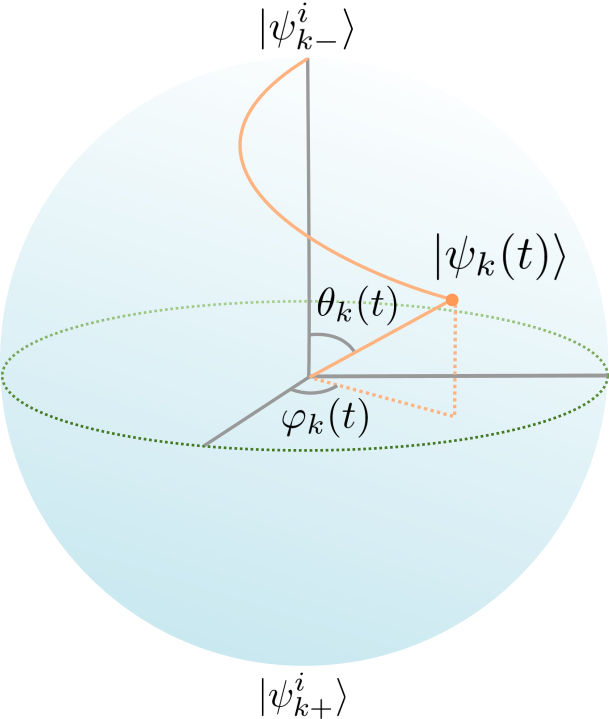}
\caption{Relative Bloch sphere representation of a single-momentum state $|\psi_k\rangle$ fully specified by the azimuthal $\phi_k$ and polar angle $\theta_k$. The north (south) pole corresponds to the lower (upper) band of the initial Hamiltonian. Real-time evolution describes a trajectory (indicated by the orange line) with time-dependent angles.}
\label{fig:BlochSphere}
\end{figure}

Generally, any state $|\psi_k\rangle$ living in a single momentum sector $k$ can be expanded in the eigenbasis of a set of Bloch states $|\psi_{k \pm} \rangle$ according to:
\be
	|\psi_k \rangle = \cos[\theta_k/2]  |\psi_{k-}\rangle + e^{i\varphi_k}\sin[\theta_k/2] |\psi_{k+}\rangle
\ee
Consequently, $|\psi_k\rangle$ admits a representation as a point on the Bloch sphere with $\theta_k \in [0,\pi]$ denoting the polar and $\varphi_k \in [0,2\pi)$ the azimuthal angle.
For an illustration, see figure~\ref{fig:BlochSphere}.

Such a representation is not only valid in equilibrium but also within dynamical processes.
When fixing some particular basis set $|\psi_{k\pm}\rangle$ we can write:
\be
	|\psi_k(t) \rangle = \cos[\theta_k(t)/2] | \psi_{k-}\rangle + e^{i\varphi_k(t)}\sin[\theta_k(t)/2] |\psi_{k+}\rangle \, ,
	\label{eq:relBlochSphere}
\ee
which can be represented as a trajectory on the Bloch sphere.
For the considered quantum quench protocol it turns out to be suitable to choose as the basis the Bloch states $|\psi_{k \pm}^i\rangle$ of the initial Hamiltonian such that at time $t=0$ the Bloch sphere representation for each momentum is located at the north pole, i.e., $\theta_k(t=0)=0$ and $|\psi_k(t=0)\rangle = |\psi_{k-}^i\rangle$.
This particular representation of the state is referred to as the 'relative Bloch sphere'~\cite{Budich2015}.
The condition for DQPTs with $\mathcal{G}_{k^\ast}(t_c)=0$ translates in the relative Bloch sphere picture to $\cos[\theta_{k^\ast}(t_c)/2] = 0$ yielding $\theta_{k^\ast}(t_c) = \pi$.
This means that $|\psi_{k^\ast}(t_c) \rangle$ is located at the south pole.

\subsection{Topological and accidental DQPTs}
\label{subsec:protectedaccidental}

We have seen that for DQPTs to occur it is sufficient for at least one mode $k^\ast$ to reach the south pole on the relative Bloch sphere.
The considerations from section~\ref{subsec:landauzener} provided a physical picture for the occurrence of this mode from Landau-Zener arguments.
It is the goal of the following to summarize the rigorous results on the occurrence of DQPTs in topological systems.
On general grounds it has been shown in~\cite{Vajna2015} under which conditions the equilibrium ground state topology necessarily imposes the existence of DQPTs.
In 1D the situation is particularly clear. 
For any quantum quench between two topologically different Hamiltonians, as characterized by their winding number, there exists at least one critical momentum $k^\ast$.
In 2D there is a richer phenomenology.
There, it can be shown that it is not sufficient to change the ground state topology as measured by the Chern number in order to be guaranteed to get DQPTs.
It is rather relevant whether the \emph{absolute} value of the Chern number differs for the two Hamiltonians appearing in the quench.
Only in these cases DQPTs necessarily appear.
Thus, for quenches in 1D between Hamiltonians with different equilibrium topological properties or in 2D with a different absolute value of the Chern number, DQPTs are robust and have therefore termed 'topological' or 'symmetry-protected'~\cite{Huang2016}.
For quenches not falling into these classes, DQPTs can still occur~\cite{Budich2015, Vajna2015, Hickey2014, Huang2016}.
In these cases they are not robust and require fine-tuning of the Hamiltonian.
These DQPTs have acquired the notion of 'accidental'.

\begin{figure*}
\centering
\includegraphics[width=1\textwidth]{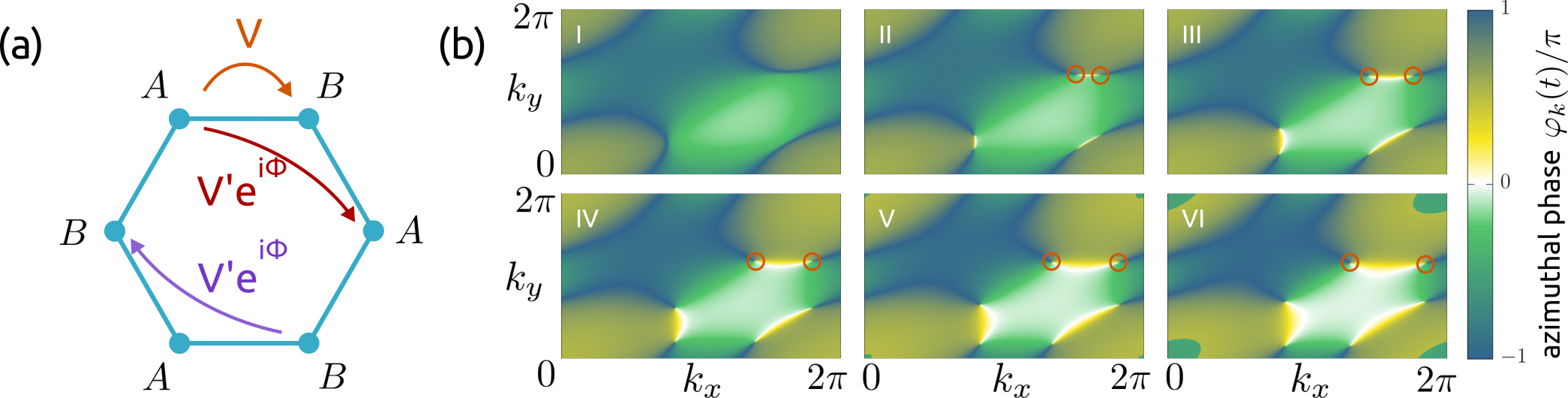}
\caption{Illustration of the vortex dynamics for a quantum quench in the Haldane model~\cite{Haldane}. {\bf (a)} Honeycomb lattice with nearest-neighbor hopping amplitude $V$ and intra-sublattice hopping $V'e^{i\Phi}$. {\bf (b)} Vortex dynamics of the azimuthal phase $\varphi_k(t)$ in the Brioullin zone (lattice spacing $a=1$) for a quantum quench across the topological phase transition in the Haldane model from initial mass $m_i/V=5$ to $m_f=0$ for fixed $V'/V=1/4$ and $\Phi=\pi/2$. The false-color plots (I to VI) show increasing times $t$ between $t=0.85/V$ to $t=1.1/V$ in steps of $\Delta t = 0.05/V$. At a time $t=0.9/V$ (II) the first vortex pairs are created. One of the pairs we track by enclosing them in the orange circles. The vortices are mobile objects moving through the Brioulling zone over time (II to VI).}
\label{fig:vortices}
\end{figure*}

\subsection{Topological defects in dynamical phase profiles}
\label{subsec:topologicaldefects}

DQPTs in topological systems come along with interesting structures in the dynamics of phase profiles.
This includes the azimuthal angle $\varphi_k(t)$ of the relative Bloch sphere, see equation~(\ref{eq:relBlochSphere}), and the so-called Pancharatnam geometric phase~\cite{Budich2015}.
The phase profile of the azimuthal angle $\varphi_k(t)$ for a 2D system has been measured experimentally, as is summarized in more detail in section~\ref{subsec:expAtoms}.
It is straightforward to see on general grounds why DQPTs have a strong impact onto the azimuthal angle $\varphi_k(t)$.
Due to unitarity of time evolution the dynamics for each momentum on the Bloch sphere describes a smooth trajectory.
When, however, the trajectory of the critical mode $k^\ast$ crosses the south pole, the state $|\psi_{k}(t) \rangle$ moves to the opposite hemisphere implying a sudden jump by $\pi$ in the azimuthal angle $\varphi_{k^\ast}(t)$.
In 2D this sudden $\pi$ phase shift leads to the creation of vortex-antivortex pairs of the full phase profile in the Brioullin zone~\cite{Flaeschner2016}.
In figure~\ref{fig:vortices}(b) this phase profile is shown for a quantum quench in a Haldane model~\cite{Haldane} which in the context of equation~(\ref{eq:defhk}) exhibits the following representation:
\begin{eqnarray}
d_k^x = V\sum_{j=1}^3 \cos(k a_j)\, , \quad d_k^y = V \sum_{j=1}^3 \sin(k a_j) \, , \\
d_k^z = m - 2 V' \sin(\Phi) \sum_{j=1}^3 \cos(k b_j) \, .
\end{eqnarray}
Here, $m$ denotes an energy offset between two $A$ and $B$ sublattices on the considered honeycomb lattice, see figure~\ref{fig:vortices}(a) for the respective real-space structure.
$V$ is the nearest-neighbor hopping amplitude and $V'e^{i\Phi}$ the complex hopping amplitude within the same sublattice.
For the definition of the lattice vectors $a_j$ and $b_j$ connecting nearest and next-to-nearest neighbor lattice sites we refer to~\cite{Haldane}.
In figure~\ref{fig:vortices}(b) we show the resulting vortex dynamics for a quantum quench from initial offset $m_i = 5V$ to final offset $m_f=0$ at a fixed $V'/V=1/4$ and $\Phi=\pi/2$ across the underlying equilibrium topological phase transition.
As one can see, there appears a critical time where suddenly pairs of vortices are created which start to move through the Brioullin zone. 
An alternative phase profile to characterize DQPTs in topological systems has made use of the concept of the Pancharatnam geometric phase~\cite{Pancharatnam56,Samuel1988} which extends the notion of Berry's phase~\cite{Berry1984dt,Simon1983yf} to general unitary evolution with non-orthogonal initial and final states.
Importantly, this phase is naturally contained in the Loschmidt amplitude.
Let us introduce a polar decomposition of $\mathcal{G}_k(t)$ at a given momentum $k$:
\be
	\mathcal{G}_k(t) = r_k(t) e^{i\phi_k(t)} \, .
\ee
The phase $\phi_k(t)$ contains a geometric part $\phi_k^P(t)$, the Pancharatnam geometric phase, by subtracting a dynamical contribution
\be
	\phi_k^P(t) = \phi_k(t) - \phi_k^\mathrm{dyn}(t) \,,
\ee
with $\phi_k^\mathrm{dyn}(t) = -\int_0^t ds \langle \psi_k(s) | H(s) | \psi_k(s)\rangle = -n_k^f t$. As the azimuthal phase for the critical momentum $k^\ast$ exhibits a $\pi$-phase slip at the DQPT, so does $\phi_k^P(t)$~\cite{Budich2015}.
In 1D particle-hole symmetry ensures that one can define an integer-valued winding number for $\phi_k^P(t)$ changing its value $\pm1$ at every DQPT~\cite{Budich2015} due to the anticipated $\pi$ phase jump.
Interestingly, from the dynamics of this winding number it is, in principle, possible to distinguish accidental from symmetry-protected DQPTs~\cite{Budich2015}.
In 2D, the $\pi$-phase slips lead to the creation or annihilation of vortex pairs in the phase profile for $\phi_k^P(t)$ in the Brioullin zone~\cite{HeylBudich2017} similar to the case of the azimuthal angle $\varphi_k(t)$ studied before.
The definition of these dynamically created or annihilated vortices for the Pancharatnam geometric phase can be also generalized to the case of mixed states~\cite{HeylBudich2017,Bhattacharya2017b}, see also section~\ref{subsec:mixedstates}.

\section{Experiments}
\label{sec:experiments}

Recently, DQPTs have been observed in two experiments performed on quantum simulators~\cite{Flaeschner2016,Jurcevic2016} that are summarized in the following.
We do not attempt to discuss experimental details, for which we refer to the respective articles, but rather focus on the main findings and implications.
While these two experiments have observed DQPTs with tailored methods, on a general level, a protocol has been recently introduced which allows to access Loschmidt amplitudes in systems of ultracold atoms~\cite{Daley2012,Pichler2013}.
Moreover, in systems where the complete quantum state can be reconstructed with full state tomography, such as in trapped ions or superconducting qubits, Loschmidt amplitudes are accessible directly.

\subsection{Trapped ions}
\label{subsec:expIons}

\begin{figure}
	\centering
	\includegraphics[width=1\columnwidth]{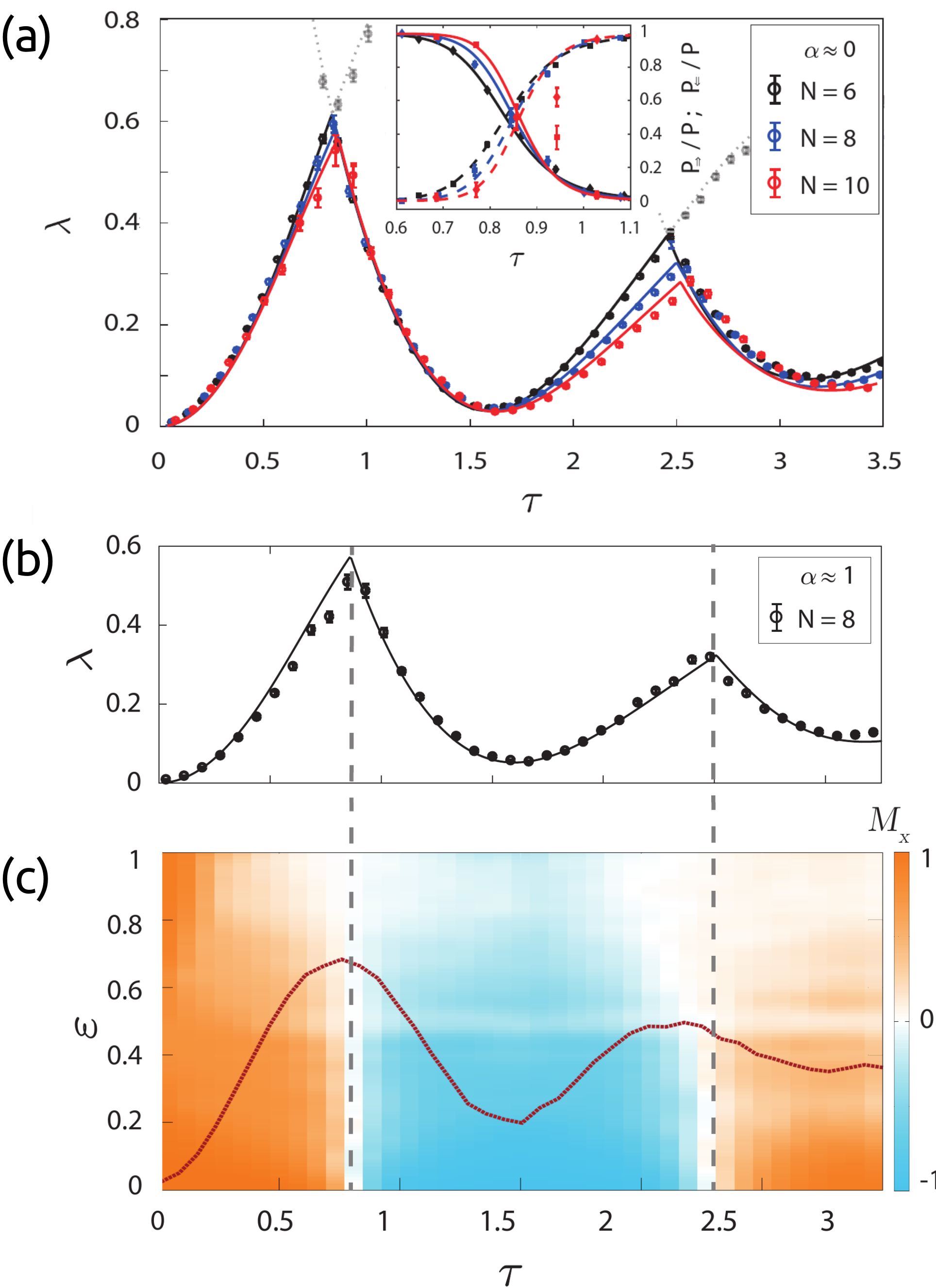}
	\caption{Dynamical quantum phase transitions in the trapped ion experiment~\cite{Jurcevic2016} {\bf (a)} Measured data for the Loschmidt echo rate function $\lambda(t)$ at $\alpha\approx 0$ and different system sizes as a function of dimensionless time $\tau=ht$ displaying clearly nonanalytic behavior. The colored data points show $\lambda(t)$, obtained via taking its dominant contribution $\lambda(t) = \min_{\eta=\uparrow,\downarrow} \lambda_\eta(t)$, whereas the grey data points refer to the respective subleading ones. {\bf (b)} Experimental result for $\lambda(t)$ at a larger interaction exponent $\alpha\approx 1$. {\bf (c)} Experimental reconstruction of the energy-resolved magnetization $\mathcal{M}(\varepsilon,t)$ displaying sharp changes  (not nonanalytic because of the finite-size quantum simulation) with $\mathcal{M}(\varepsilon=0,t)$ changing sign along $\varepsilon=0$. This sharp feature fades out to $\varepsilon>0$ eventually crossing the mean energy density $\varepsilon_\mathrm{av}(t)$, included as the red line, where local observables attain their dominant contribution.}
	\label{fig:expIons}
\end{figure}

Systems of trapped ions can synthesize the dynamics of transverse-field Ising models of the form~\cite{Lanyon2011, Britton2012, Jurcevic2014, Richerme2014, Smith2016, Jurcevic2016}
\be
H = -\sum_{l>m} J_{lm} \sigma_l^z \sigma_m^z - h \sum_{l=1}^N \sigma_l^x \, .
\label{eq:expIonsHamLRIsing}
\ee
Here, $\sigma_l^\alpha$ with $\alpha=x,y,z$ denote the Pauli operators on lattice site $l=1,\dots,N$ with $N$ the total number of spins.
The coupling $J_{lm}$ is approximately of long-ranged form~\cite{Islam2013}
\be
J_{lm} \sim \frac{1}{|l-m|^\alpha}, \quad \mathrm{ for } \,\, |l-m| \gg 1 \, ,
\ee
with a tunable interaction exponent $\alpha$ from $\alpha=0$ up to $\alpha=3$.

In the trapped ion experiment on DQPTs~\cite{Jurcevic2016} a quantum quench across the ferromagnetic to paramagnetic equilibrium phase transition has been realized -- a situation where generically DQPTs are expected.
Initially, the system is prepared in the fully polarized state
\be
|\psi_0\rangle = |\uparrow \rangle = |\uparrow \dots \uparrow\rangle \, ,
\ee
which is one of the two ground states of the Hamiltonian in equation~(\ref{eq:expIonsHamLRIsing}) for vanishing transverse field $h=0$.
After this stage of preparation the subsequent dynamics of the system is driven by $H$ at a transverse field $h$ sufficiently large such that in equilibrium the system would reside in a paramagnetic phase. 
Since the system is in a symmetry-broken phase at zero temperature for the initial Hamiltonian at $h=0$, we use the generalization $P(t)$ from equation~(\ref{eq:groundStateProb}) for the Loschmidt echo:
\be
P(t) =  P_\uparrow(t) + P_\downarrow(t) \, ,
\ee
which is the full probability to return to the ground state manifold at a time $t$, with
\be
P_\uparrow(t) = \big| \langle \uparrow | \psi(t) \rangle \big|^2, \,\, P_\downarrow(t) = \big| \langle \downarrow | \psi(t) \rangle \big|^2 \, .
\ee
As discussed in section~\ref{subsec:loschmidt}, the wave function overlaps we consider all have an exponential dependence on system size $N$ such that:
\be
P_\eta(t) = e^{-N\lambda_\eta(t)}, \,\,\, \eta = \uparrow,\downarrow \, ,
\ee
with $\lambda_\eta(t)$ intensive functions independent of $N$ in the thermodynamic limit.
This property has important implications for the rate function $\lambda_N(t) = -N^{-1} \log[P(t)]$ of the full $P(t)$.
In particular, $P(t)$ for $N\to\infty$ is always dominated by one of the two contributions $P_\uparrow(t)$ or $P_\downarrow(t)$ such that~\cite{Heyl2014,Zunkovic2016,Jurcevic2016}
\be
\lambda(t) = \lim_{N\to\infty} \lambda_N(t) =  \min_{\eta=\uparrow,\downarrow} \lambda_\eta(t) \,.
\label{eq:lambdaMinimum}
\ee
This is the central tool to predict DQPTs from the experiment where the $P_\eta(t)$, $\eta=\uparrow,\downarrow$, can measured individually.
Let us define a function $\tilde \lambda_N(t) = \min_{\eta=\uparrow,\downarrow} \lambda_\eta(t)$, which  for a finite system is different from $\lambda_N(t)$ but converges to the same $\lambda(t)$ in the thermodynamic limit.
And let us suppose, that one can reach system sizes where the individual $\lambda_\eta(t)$ can be considered as converged with negligible finite-size corrections.
Then, we obtain that $\tilde{\lambda}_N(t) \approx \lambda(t)$ and the finite-size data can already be used to predict the behavior in the thermodynamic limit.
Of course, this procedure implies an additional theoretical input to the experiment.
The measured data for the individual rate functions $\lambda_\eta(t)$ is shown in figure~\ref{fig:expIons}(a) and \ref{fig:expIons}(b).
As one can see, the $\lambda_\eta(t)$ have already almost converged at least for times within the first half of the shown data.
Consequently, one can use the $\lambda_\eta$'s to construct $\tilde \lambda_N(t)$ which due to (almost) convergence with $N$ is equivalent to $\lambda(t)$, so that in the following we don't have to distinguish $\tilde{\lambda}_N(t)$ from $\lambda(t)$.
Since there appear points in time where the two $\lambda_\eta(t)$'s intersect, $\lambda(t)$ develops a kink in the thermodynamic limit $N\to\infty$.
With this theoretical input, the experimental data implies nonanalytic real-time dynamics in particular.
Without this theoretical input of the minimum construction, one could have studied $\lambda_N(t)$ instead.
In contrast to $\tilde{\lambda}_N(t)$, $\lambda_N(t)$ is a smooth function for a finite system, which, however, becomes sharper at the critical time for increasing system size eventually leading to the nonanalytic behavior of $\tilde{\lambda}_N(t)$ in the thermodynamic limit, as also discussed in reference~\cite{Jurcevic2016}.

In the experiment not only DQPTs have been observed but also the relation to other observables has been systematically studied.
This includes in particular a quantitative approach to the analogy between DQPTs and conventional equilibrium QPTs discussed on general grounds in section~\ref{subsec:QPT}.
For the initial Ising Hamiltonian $H_0$ at vanishing field $h=0$, we have that $[H_0,\mathcal{M}]=0$ with $\mathcal{M}=N^{-1}\sum_l \sigma_l^z$ the magnetization.
Consequently, $H_0$ and $\mathcal{M}$ can be measured simultaneously which can be used to define an energy-resolved magnetization $\mathcal{M}(\varepsilon,t)$ at a given time $t$ with $\varepsilon$ denoting the energy density~\cite{Heyl2014,Jurcevic2016}.
Due to the measurement capabilities in trapped ions it is possible to also experimentally access this quantity which is shown in figure~\ref{fig:expIons}(c).
As one can clearly see, there appears a temporal analog of a quantum critical region in the energy density-time plane which is controlled by the DQPT occuring at $\varepsilon=0$ (upon choosing the zero of energy accordingly) at a critical time $t_c$.
Moreover, the experiment has studied entanglement production and observed a close connection, see also section~\ref{subsec:entanglement} for a more detailed discussion of entanglement properties at DQPTs.

\subsection{Ultra-cold atoms in optical lattices}
\label{subsec:expAtoms}

\begin{figure}
	\centering
	\includegraphics[width=1\columnwidth]{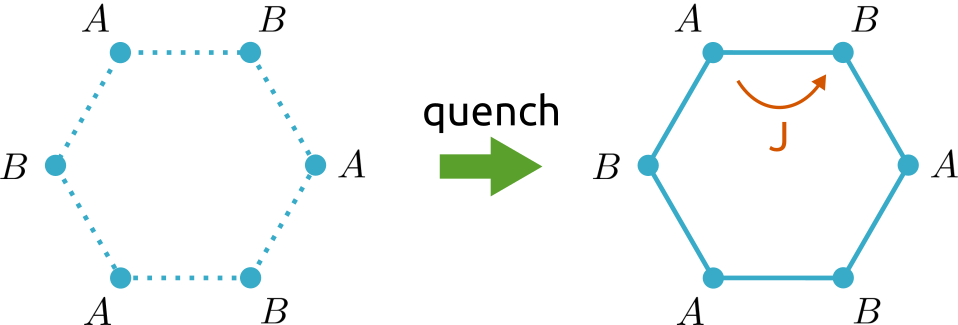}
	\caption{Illustration for the setup used in the observation of dynamical quantum phase transitions in the ultra-cold atom experiment~\cite{Flaeschner2016}.}
	\label{fig:expAtoms}
\end{figure}

The ultra-cold atom experiment on DQPTs~\cite{Flaeschner2016} has observed  dynamical \emph{topological} quantum phase transitions.
While the general theory of such DQPTs has already been discussed in section~\ref{sec:DPTP}, it is the goal of the present section to outline and discuss experimental aspects.

This experiment has synthesized a quantum quench in a system of noninteracting fermionic degrees of freedom on a hexagonal lattice, see figure~\ref{fig:expAtoms} for an illustration.
For the initial preparation, a large energetic offset between two sublattices $A$ and $B$ has been imposed, such that the system realizes a simple insulating ground state of a two-band system at half filling to a very good approximation with the particles localized on the lattice sites of the $B$ sublattice.
Then, at time $t=0$ lattice shaking Floquet techniques are used to suddenly couple the two sublattices -- effectively switching on the hopping -- which realizes the quantum quench and induces nonequilibrium dynamics in the system.

As discussed already in section~\ref{sec:DPTP}, the dynamics in such topological system can be decomposed into contributions from all the (crystal) momenta $k$ of the Brioullin zone.
Moreover, for each momentum $k$ the wave function $|\psi_k(t)\rangle$ for such a two-band model reduces to an effective two-level system admitting a representation on the Bloch sphere with two associated angles: the polar angle $\theta_k(t)$ and azimuthal angle $\varphi_k(t)$.

Using full-state tomography techniques for two-band noninteracting fermionic systems~\cite{Hauke2014,Flaeschner2016a} the experiment obtained access to both of these angles.
Of particular interest in the context of DQPTs is the azimuthal $\varphi_k(t)$, whose dynamics has already been anticipated in section~\ref{subsec:topologicaldefects}.
Monitoring the dynamics of $\varphi_k(t)$ in the Brioullin zone one can observe that there appear points in time where pairs of vortices are created or annihilated.
Importantly, such a sudden creation or annihilation of vortex-antivortex pairs is in a one-to-one correspondence with an underlying DQPT independent of the model details~\cite{Flaeschner2016} meaning that the associated number of dynamical vortices can change if and only if the system experiences a DQPT as long as it can be considered as noninteracting.
Consequently, tracking this vortex number over time is equivalent to tracking DQPTs.
This is of particular importance since the experiment only provides access to a discrete set of points in time as it is realized as a Floquet system and is therefore only monitored stroboscopically.
While from the full-state tomography also, in principle, the Loschmidt echo rate function $\lambda(t)$ can be reconstructed~\cite{Flaeschner2016}, it is not possible to uniquely identify nonanalytic real-time structures from a fixed time grid.
In this context the dynamical vortex number is appealing due to its quantized nature in that a change in this number can only happen nonanalytically.
Even more, the vortex number might not only be viewed as a way to detect DQPTs but can also be interpreted as a dynamical order parameter for the DQPTs in this model~\cite{Flaeschner2016}.

\section{Characteristic properties}
\label{sec:generalPrinciples}

The nonanalyticities at DQPTs and the formal similarity of Loschmidt amplitudes to equilibrium partition functions suggest a close connection between DQPTs and conventional phase transitions.
Equilibrium transitions, however, entail many further key properties  beyond the mere nonanalytic character of thermodynamic potentials.
It is one of the major challenges for the theory of DQPTs to identify proper extensions of such characteristic principles to the far-from-equilibrium regime.
It is the main purpose of this section to summarize and discuss results for the theory of DQPTs that address such fundamental questions.
It is important to emphasize, however, that the current understanding rather represents a first step towards a comprehensive theory for nonequilibrium phase transitions.
The summarized results are supposed to be seen as a starting point for further investigations towards this major goal.

\subsection{Scaling and universality}
\label{subsec:universality}

Let us start by discussing to which extent the concepts of scaling and universality, which in equilibrium are caused by a divergent correlation length, can be applied to DQPTs.
While a general understanding has not yet been reached, for the Ising model these concepts can be extended to the dynamical regime~\cite{Heyl2015dq}.
It is the goal of the following section to summarize the main idea and to discuss the implications.

Consider the transverse-field Ising model
\be
      H(h) = - J \sum_{\langle l m \rangle } \sigma_l^z \sigma_m^z - h \sum_{l=1}^L \sigma_l^x,
\ee
with the Pauli-matrices $\sigma_l^\alpha$, $\alpha=x,y,z$, and $l=1,\dots,L$ where $L$ is total number of lattice sites.
Here, $\langle \dots \rangle$ denotes a summation over nearest-neighboring lattice sites.
At the moment, let us not restrict to a particular dimension or graph.

Let us now focus on a particular quantum quench from $J/h=0$ to $h/J=0$.
While this quench is very specific, it represents a fruitful starting point for approaching the problem on a general level. 
Within this nonequilibrium scenario, the system is initialized in the fully polarized state along transverse-field direction:
\be
    |\psi_0\rangle = |+\rangle = \bigotimes_l |+\rangle_l,
\ee
where $\sigma_l^x |+\rangle_l = |+\rangle_l$, and the time evolution is then governed by the final Hamiltonian $H = - J \sum_{\langle l m \rangle } \sigma_l^z \sigma_m^z$.

It is the central observation, that in this case the Loschmidt amplitude can be mapped onto a conventional partition function
\be
    \mathcal{G}(t) = Z(K) =  \frac{1}{2^L} \mathrm{Tr} e^{\mathcal{H}(K)} \, ,
\ee
of a classical Ising model with Hamiltonian
\be
    \mathcal{H}(K) = K \sum_{\langle lm\rangle} \sigma_l^z \sigma_m^z, \quad K = iJt \, .
\ee
The only difference to the equilibrium case is that the coupling $K \in \mathbb{C}$ appearing in $\mathcal{H}(K)$ is complex.
As discussed in section~\ref{subsec:fisherZeros}, Loschmidt amplitudes can always be formally understood as conventional \emph{boundary} partition functions at complex parameters.
The key property of the particular considered quantum quench is that the boundary conditions can be fully absorbed into the \emph{bulk}.

The equivalence between $\mathcal{G}(t)$ and $Z(K)$ can be seen straightforwardly when recognizing that the initial state $|+\rangle = 2^{-N/2} \sum_s |s\rangle$ is an equally weighted superposition of all spin configurations.
Because the Hamiltonian $H$ governing the time evolution is diagonal in the spin basis, i.e., $H |s\rangle = H(s) |s \rangle$ with $H(s)$ the respective eigenvalue, we have that
\be
	\mathcal{G}(t) = \sum_{ss'} \langle s' | e^{-itH(s)} | s\rangle = \sum_s e^{-itH(s)} = \mathrm{Tr} e^{\mathcal{H}(K)} \, ,
\ee
which gives the desired relation.
The major advantage of this construction is that now results and strategies known for the equilibrium case can be applied to Loschmidt amplitudes and thus DQPTs.
This is particularly interesting in the 1D and 2D cases as will be discussed now.
Let us start with the 1D Ising chain where it is possible to construct an exact renormalization group (RG) transformation allowing for the identification of the exact RG fixed points.
Specifically, it is possible to apply conventional decimation RG procedures~\cite{Nelson1975zz} to the complex partition function of interest here~\cite{Heyl2015dq}.
Eliminating every second lattice site, one obtains the following exact recursion relation for change of the couplings within one RG step:
\be
	\tanh(K') = \tanh^2(K) \, ,
\label{eq:decimationRGEquation}
\ee
which precisely matches the equilibrium case with the only difference that in the present dynamical scenario the effective coupling $K\in\mathbb{C}$ is in general complex.
This leads to the immediate question whether the extension of the coupling $K$ into the complex plane can lead to new fixed points.
It turns out that only the conventional stable $K^*=0$ (infinite temperature) and unstable $K^\ast=\infty$ (zero temperature) fixed points can be reached. 

\begin{figure}
\centering
\includegraphics[width=\columnwidth]{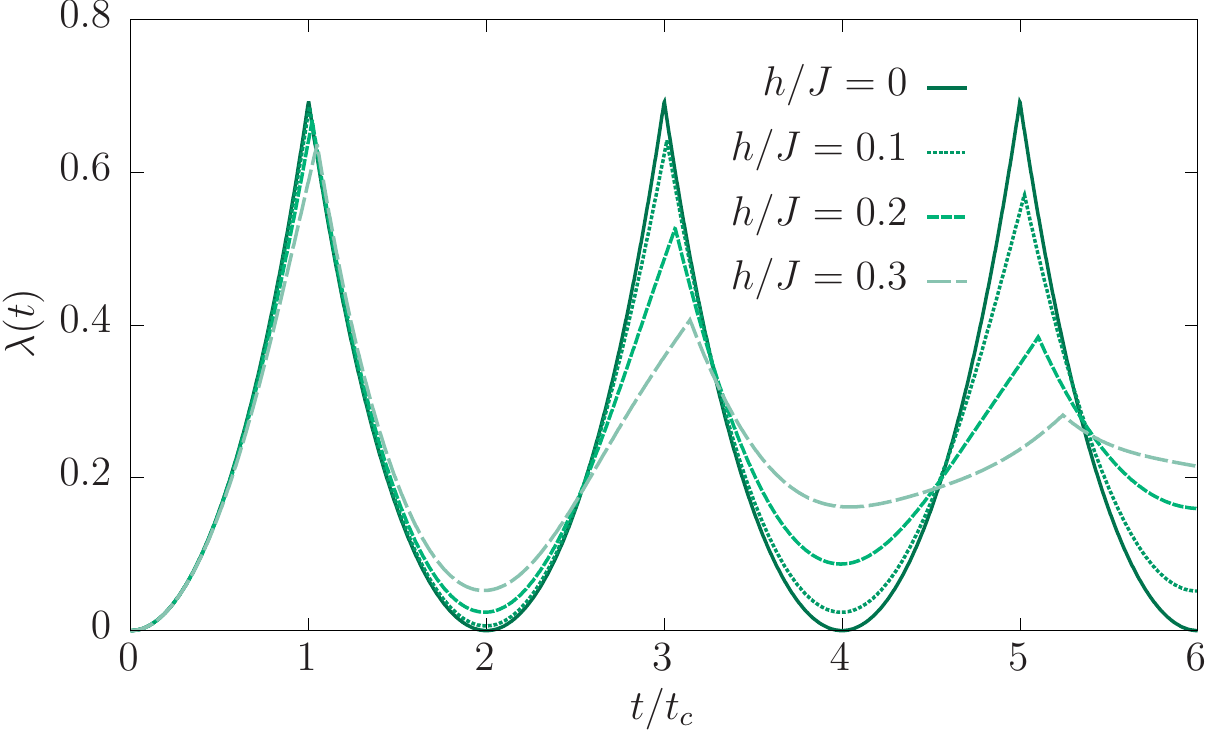}
\caption{DQPTs in the Loschmidt echo rate function $\lambda(t)$ for quantum quenches in the transverse-field Ising chain for varying final fields $h$ starting from initial fully polarized states in the transverse field direction~\cite{Heyl2015dq}. At the critical times $\lambda(t)$ exhibits a kink as predicted by scaling theory, see equation~(\ref{eq:isingKinkScaling}). While quantitatively for $h>0$ the rate function $\lambda(t)$ shows deviations upon varying $h$, the universal properties of the DQPT in form of the kinks remain unaltered in agreement with the RG prediction.}
\label{fig:dqptIsingExtremeQuench}
\end{figure}

To which fixed point are the DQPTs associated with? Taking the critical coupling $K_c = i\pi/4$, where the system exhibits a DQPT, and applying the RG recursion relation in equation~(\ref{eq:decimationRGEquation}) one obtains that $K_c \mapsto K^\ast=\infty$ maps into the \emph{unstable zero-temperature} fixed point of the equilibrium Ising model.
This directly implies that these DQPTs obey scaling and universality.
As a consequence, one can immediately obtain the universal scaling form of the singular contribution $g_s(t)$ to the dynamical free energy density $g(t)$ as:~\cite{Heyl2015dq}
\be
	g_s(t) \sim |\tau|, \quad \tau = \frac{t-t_c}{t_c},
	\label{eq:isingKinkScaling}
\ee
with $\tau$ denoting the dimensionless distance from the critical time $t_c$.
The temporal kink in $g(t)$ obtained by the scaling analysis matches precisely the result from the full exact solution, see figure~\ref{fig:dqptIsingExtremeQuench}.

What one can gain from the relation between the DQPT and an unstable fixed point, for example, is that it is now straightforward to systematically study the influence of perturbations to the model.
In particular, perturbations irrelevant in the RG sense leave the fixed point unchanged which implies also a certain robustness of DQPTs. 
For example, adding a next-to-nearest spin coupling is always irrelevant under the decimation RG, which  holds independent of the associated coupling strength~\cite{Nelson1975zz}.
Moreover, one can also start studying the influence of a transverse field in the final Hamiltonian.
Within a perturbative treatment this leads to an effective classical description with an effective Ising model including weak irrelevant next-to-nearest neighbor interactions~\cite{Heyl2015dq}.
Interestingly there is the possibility that the relevance of a perturbation, although appearing potentially with a weak coupling in the Hamiltonian, might depend on time because the effective couplings appearing in the decimation RG implicitely exhibit a time-dependence.
Let us continue by studying the 2D Ising model on a square lattice where the identification of the Loschmidt amplitude with a classical partition function is again possible.
Although in this case no exact RG transformation can be formulated, the dynamics for the Loschmidt amplitude can still be accessed extending the solution for the partition function of the 2D Ising model~\cite{Onsager1944kb,Kaufman1949pp,Schultz1964qt} to complex couplings~\cite{Heyl2015dq}.
One finds that this system also exhibits a DQPT.
The singular contribution $g_s(t)$ to the Loschmidt amplitude rate function displays a nonanalytic behavior:
\be
	g_s(\tau) \sim \tau^2 \log( |\tau|).
\ee
Remarkably, this matches precisely the critical behavior of the free energy density at the thermal critical point of the 2D Ising model suggesting that scaling and universality also hold for that case.

Currently, no other examples except the discussed models are known for which scaling and universality at DQPTs have been established.
Overall, the DQPTs discussed in this section appear to exhibit the scaling which is not associated with the underlying quantum equilibrium phase transition, as also observed for a 1D quantum Potts chain~\cite{Karrasch2017}, but rather to the classical one.
Investigating to which extent universality and scaling generalize to the other DQPTs and whether also genuine nonequilibrium fixed points can appear, that are not related to equilibrium criticality, is a pertinent task for future work on the theory of DQPTs.

\subsection{Robustness}
\label{sec:robustness}

The robustness of DQPTs has been studied for many models~\cite{Pollmann2010dv, Karrasch2013, Kriel2014, Heyl2015dq, Sharma2015}.
For DQPTs exhibiting scaling and universality as discussed in the previous section, robustness against a large class of perturbations is guaranteed.
Whenever the perturbation is weak in the sense of the constructed RG, the structure of the nonanalytic behavior is unchanged while only nonuniversal aspects such as the critical time of the DQPT might be shifted.
An example of such an RG irrelevant perturbation has been provided in Ref.~\cite{Heyl2015dq}.
Upon adding a transverse field to the Ising Hamiltonian discussed in the previous section, the Loschmidt amplitude becomes equivalent to a classical Ising model including next-to-nearest neighbor interactions.
These longer-ranged couplings are irrelevant and vanish under the RG transformation such that the fixed point Hamiltonian is described again by an Ising model without a transverse field at, however, renormalized couplings.
Consequently, the DQPTs are robust in the sense that the nonanalytic structure does not change and the sole influence of the transverse field is to shift the critical time.

As a further supporting argument for the robustness of DQPTs under symmetry-preserving perturbations one can resort to the formal similarity of Loschmidt amplitudes to complex partition functions, which has already been discussed in section~\ref{subsec:fisherZeros}.
As for equilibrium transitions, DQPTs are linked in a one-to-one correspondence with complex partition function zeros.
For equilibrium partition functions it is well known that the structures formed by these zeros in the complex plane are typically robust under weak symmetry-preserving perturbations, which is a different way of seeing the stability of phase transitions without resorting to an RG analysis.
These structures might deform but do not immediately melt under the addition of a weak perturbation, which is equivalent to changing the critical value of the control parameter but retaining the structure of the transition.
Assuming that complex partition function zeros show the same properties also in the whole complex plane, also DQPTs are then expected to be robust.
This, however, should not been seen as a proof, but rather as a general physical argument.

For concrete models the stability of DQPTs has been studied both using numerical and analytical approaches~\cite{Pollmann2010dv, Karrasch2013, Kriel2014, Heyl2015dq, Sharma2015}.
Specifically, in these works variants of the transverse-field Ising chain in 1D have been investigated under the inclusion of different perturbations.
When adding a symmetry-preserving next-to-nearest neighbor interaction to the model, which has been shown on the basis of both numerical simulations using the time-dependent density-matrix renormalization group approach~\cite{Karrasch2013} as well an analytical approach using the flow equation method~\cite{Kriel2014}.
The influence of a magnetic field in the ordering direction of the Ising chain, which constitutes a symmetry-breaking perturbation, can lead to a smearing of the DQPT for a parameter sweep~\cite{Pollmann2010dv}
Adding such a perturbation for a quantum quench, however, it can occur that DQPTs still exist~\cite{Karrasch2013,Sharma2015}.
Understanding the difference between slow and fast perturbations in this context remains an open question~\cite{Karrasch2013}.

All the summarized examples study the robustness of DQPTs on short to intermediate time scales.
A different question is how the DQPTs are influenced in the long-time limit where it is known from the context of quantum thermalization~\cite{Polkovnikov2011kx} that already vanishingly weak perturbation can have a strong impact onto the dynamics.
Specifically, an integrable model can be turned into an ergodic one which implies a drastic change in the asymptotic long-time steady state from a nonthermal to a thermal one.
In this light, the robustness of DQPTs might depend on time.
This question, however, has not yet been studied, but might provide an interesting connection with the field quantum thermalization.
Let us, however, emphasize that DQPTs are not relying on integrability.
For example, DQPTs have also been found for genuinely interacting models such as the Hubbard model in high dimensions~\cite{Canovi2014fo}, which is known to exhibit quantum thermalization~\cite{Eckstein2009wj}.

\subsection{Dynamical order parameters}
\label{sec:dop}

Order parameters are central for the characterization of phase transitions in equilibrium.
Therefore, it is an important question whether this concept can be extended to the considered dynamical regime.
Beyond providing a further element towards putting DQPTs on comparable footing to equilibrium transitions, dynamical order parameters might help in understanding the respective DQPT by, for example, identifying the nature of the two 'dynamical phases' separated by the DQPT. 

Dynamical order parameters have been formulated and identified for DQPTs in 1D and 2D topological systems~\cite{Budich2015,Flaeschner2016,Bhattacharya2017,Bhattacharya2017b,HeylBudich2017}.
Most notably, the 2D case has also been measured experimentally recently~\cite{Flaeschner2016}.
All the proposed dynamical topological order parameters share the same property in that they assign quantum numbers to phase profiles discussed in section~\ref{subsec:topologicaldefects}.
Importantly, these quantized integers necessarily jump at DQPTs.
For 1D systems~\cite{Budich2015,Sharma2016}, or along closed 1D paths in the 2D Brioullin zone~\cite{Bhattacharya2017}, these topological order parameters are winding numbers for the Pancharatnam geometric phase as discussed already in section~\ref{subsec:topologicaldefects}.
This winding number is capable to provide insights into the underlying ground state topology of the respective quantum many-body system although during the nonequilibrium process the system by no means is close to its ground state.
Specifically, these winding numbers can, in principle, distinguish topologically protected DQPTs from accidental ones and thus can be used to detect whether a topological quantum phase transition has been crossed with a change in the underlying equilibrium topology~\cite{Budich2015}.
In 2D, another dynamical order parameter can be constructed by measuring the number of vortices created in phase profiles across the whole Brioullin zone~\cite{Flaeschner2016,HeylBudich2017}.
Here, it is possible to choose both the Panchartnam geometric phase~\cite{HeylBudich2017} or the azimuthal phase~\cite{Flaeschner2016} of the associated relative Bloch sphere, see section~\ref{subsec:relativeBlochSphere}.
Notice that the dynamical topological order parameter using the Pancharatnam geometric phase can be generalized also for mixed initial states~\cite{HeylBudich2017} as will be discussed in more detail in section~\ref{subsec:mixedstates}.

In which sense DQPTs can also be characterized by \emph{local} dynamical order parameters is not yet known.
Certainly, this cannot be associated with long-ranged correlations in the conventional sense, because causality in terms of Lieb-Robinson bounds~\cite{lieb1972,Hastings2006,Nachtergaele2006} prevents the buildup of such long-ranged correlations within a finite time which is where DQPTs occur.
This, however, does not rule out alternative notions of a divergent correlation length which one can imagine within the interpretation of DQPTs as dynamical analogs of equilibrium quantum phase transitions, see section~\ref{subsec:QPT}. 

\subsection{Signatures in other quantities}
\label{sec:macroscopics}

Equilibrium phase transitions not only manifest in a nonanalytic free energy but also in other observables such as the order parameter, susceptibilities, or entanglement to name just a few.
In particular, it is possible to directly infer measurable quantities such as the specific heat from the free energy by taking derivatives.
At this point a difference between DQPTs and conventional transitions becomes apparent.
It is not possible to obtain other measurable quantities from Loschmidt amplitudes in a similar way.
How can DQPTs then be related to other quantities?

4In section~\ref{subsec:QPT} a general argument has been presented how DQPTs are, in principle, capable of controlling the dynamics of other quantities.
Within this argument, the DQPT is interpreted as a dynamical counterpart to a quantum phase transition in equilibrium.
A DQPT can then control the dynamics of observables whenever there exists a dynamical complement to a quantum critical region which for specific systems has been established both theoretically~\cite{Heyl2014} as well as experimentally~\cite{Jurcevic2016}.
This perspective onto DQPTs has been successful in explaining the observed signatures in certain observables of systems with symmetry-broken phases~\cite{Heyl2013a, Heyl2014, Zunkovic2016, Jurcevic2016, Weidinger2017}.
Since we are dealing with situations where ground states are not unique, it is important to specify the generalization of the Loschmidt amplitude which is also not unique, see the discussion in section~\ref{subsec:loschmidt}.
In the subsequent discussion, we follow the works in Ref.~\cite{Heyl2013a, Heyl2014, Zunkovic2016, Jurcevic2016, Weidinger2017} and define the generalization as in equation~(\ref{eq:groundStateProb}).
If alternatively, one would still choose the expression according to equation~(\ref{eq:defG}), DQPTs can also occur~\cite{Zunkovic2016a,Halimeh2016,Halimeh2017pa,Homrighausen2017}.
For this choice the connection to other observables, however, is not known which represents an interesting aspect to study in the future.

\begin{figure}
	\centering
	\includegraphics[width=1\columnwidth]{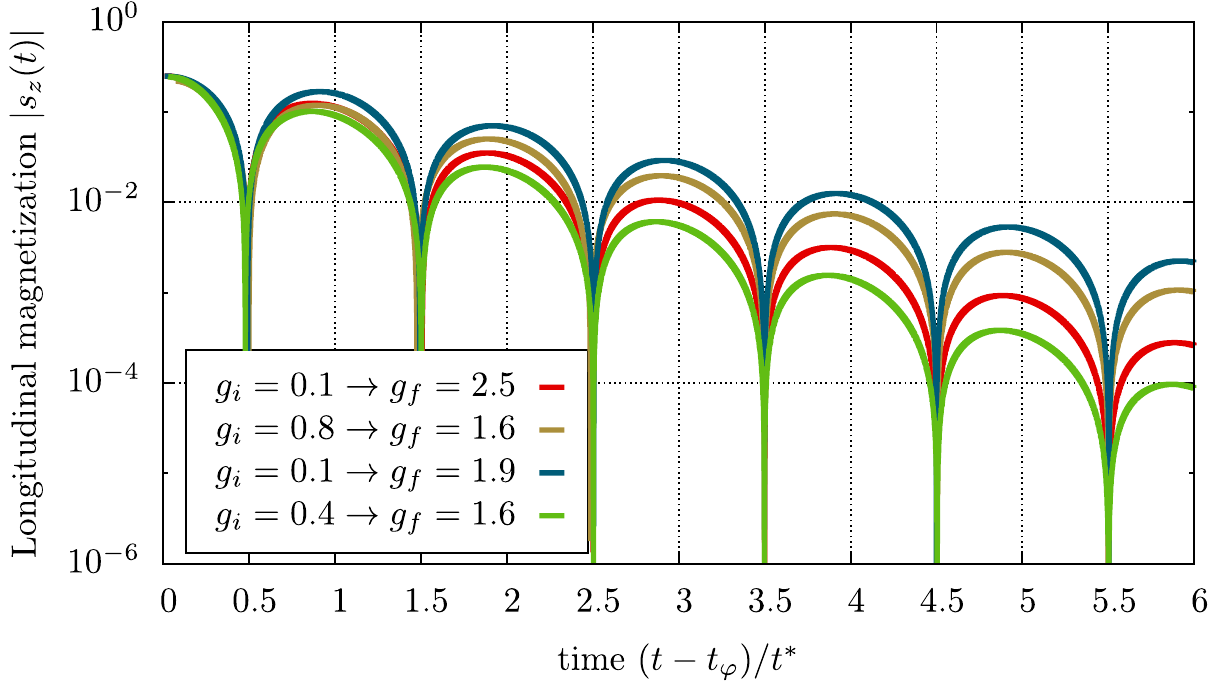}
	\caption{Decay of the the longitudinal magnetization in the transverse-field Ising chain for various initial ($g_i$) and final ($g_f$) fields~\cite{Heyl2013a}. When rescaling the time axis by $t^\ast$, which is the time of the first DQPT in the dynamics for a given parameter set, at a constant offset $t_\varphi$ one can identify the periodicity of the oscillatory decay with the periodicity at which DQPTs appear.}
	\label{fig:magdqpt}
\end{figure}

Let us now consider a system, which is prepared initially in a symmetry-broken ground state with a nonzero value of the order parameter.
Monitoring the dynamics after a quantum quench in such systems, it has been found on a rather general level that a DQPT is typically accompanied with a zero of the order parameter and therefore a periodic sequence of DQPTs with an oscillatory decay of the order parameter~\cite{Heyl2013a, Heyl2015dq, Zunkovic2016, Weidinger2017}.
In a quench in the transverse-field Ising chain, for example, the frequency of the oscillatory decay in the longitudinal magnetization matches exactly the periodicity of the DQPTs in the model independent  of the details of the quench, see figure~\ref{fig:magdqpt} for the data shown in Ref.~\cite{Heyl2013a}.
Importantly, the associated time scale $t^\ast$ is an emergent nonequilibrium time scale without equilibrium counterpart~\cite{Calabrese2012, Heyl2013a} which within the present knowledge only appears in the DQPTs and the anticipated order parameter dynamics providing a strong evidence for a connection between the two quantities.
The underlying mechanism for this connection follows a reasoning already discussed in section~\ref{subsec:expIons} on the recent trapped ion experiment and will now be outlined in more detail.
For the Ising model the full return probability to the ground state manifold $P(t)$ as defined in equation~\ref{eq:groundStateProb} is given by
\be
	P(t) = P_+(t) + P_-(t), \quad P_\eta(t) = | \langle \eta | \psi_0(t) \rangle |^2 \, ,
\ee
with $\eta = \pm$ and $|\eta \rangle$ denotes one of the two ground states with positive ($\eta = +$) and negative ($\eta=-$) magnetization.
As for the Loschmidt echo also the individual probabilities $P_\eta=\exp[-N\lambda_\eta(t)]$ exhibit a large-deviation scaling where the rate functions $\lambda_\eta(t)$ are intensive independent of the number of degrees of freedom $N$ in the thermodynamic limit.
Due to this exponential dependence on $N$ we have that $P(t)$ is always dominated by either $P_+(t)$ or $P_-(t)$ such that $\lambda(t) = -N^{-1}\log[P(t)] = \min_\eta \lambda_\eta(t)$ for $N\to\infty$.
Consequently, when the two rate functions $\lambda_\eta(t)$ cross a DQPT in $\lambda(t)$ occurs in the form of a kink, as discussed in section~\ref{subsec:expIons}.
Let us emphasize, that this kink is not a result of an artificial construction but rather carries a physical meaning which allows to connect the DQPT to other observables.
At the crossing point of the DQPT, the symmetry in the ground state probability $P(t)$ is restored, which has been initially broken explicitly by the initial condition.
It is generically found that this symmetry restoration is not just restricted to the ground state manifold but rather extends to nonzero energy densities, see for example figure~\ref{fig:expIons}.
As a consequence, the symmetry is also restored for observables which implies a vanishing value for the order parameter.
While in this discussion we have restricted to a discrete broken symmetry, the generalization to continuous symmetries is straightforward yielding a similar connection between order parameter dynamics and DQPTs~\cite{Weidinger2017}

An analogous relation between order parameter dynamics and DQPTs has been found for certain 1D topological systems with phases characterized by string order parameters~\cite{Budich2015}.
For some models these observations can be traced back to the previously discussed case of symmetry broken phases.
The Kitaev chain for particular parameter sets, for example, is equivalent to a transverse-field Ising model and the string order parameter maps onto the respective conventional order parameter correlations.
Importantly, however, the ground states of the Kitaev chain and transverse-field Ising model are different~\cite{Calabrese2012} implying that the full ground-state return probability $P(t)$ reduces to a conventional Loschmidt amplitude~\cite{Heyl2013a}.
Despite of these subtleties the main phenomenology has been found to be unchanged~\cite{Budich2015}.
When preparing the system initially in a topological phase characterized by a nonzero string order parameter, the dynamics of this string order parameter after a quantum is linked to DQPTs in the model.
Whenever the system experiences DQPTs, which due to the integrability of the model appear periodically, see equation, the string order parameter exhibits an oscillatory decay.
Again the time scale associated with these oscillations is an emergent nonequilibrium scale without equilibrium counterpart and coincides with the periodicity of DQPTs.
The discussed periodic appearance of DQPTs strongly relies on the integrable nature of the considered models.
It is therefore a natural question how integrability-breaking perturbations might influence both DQPTs and order parameter dynamics.
For strong integrability-breaking perturbations it has been found that the connection between DQPTs and order parameter dynamics can become more subtle~\cite{Karrasch2013} since additional DQPTs on longer times can appear which don't become manifest in the order parameter.
While a possible explanation for this phenomenon si that , it has not yet been resolved
A further aspect in the context of the influence of (weak) perturbations potentially inducing quantum thermalization concerns the behavior on long time scales, as discussed already in section~\ref{sec:robustness}.
This is not known yet, but represents an interesting an important aspect worthwhile to study in the future.
The previous examples summarize the relation between DQPTs and other observables for quantum quenches out of a symmetry-broken or topological phase.
For quenches in the other direction, meaning from a symmetric phase to a parameter set, where the Hamiltonian exhibits symmetry breaking in the ground state, other observables have been found which connect to the DQPTs observed in these models.
Since the initial state is symmetric and the Hamiltonian by definition conserves the symmetry, the order parameter has to vanish throughout the dynamics.
However, it has been observed that the respective order parameter correlations can exhibit signatures of the underlying DQPTs for transverse-field Ising models~\cite{Heyl2015dq, Schmitt2017}.
Specifically, they become maximal at a DQPT with a functional dependence as a function of the temporal distance to the DQPT characteristic of the nature of the DQPT in the following sense.
For the DQPTs in the transverse-field Ising model for such quenches, it has been found that the DQPTs in the 1D (2D) system are in the same universality class as the critical points of the classical 1D (2D) Ising chain.
Let us point out that, in general, identifying observables, whose dynamics is sensitive to an underlying DQPT, can be difficult because also promising candidates might not show apparent signatures. One example in this direction has been provided in a system of interacting bosons exhibiting a superfluid to Mott insulator transition in the ground state. For a quantum quench from an initial superfluid state to large interactions it has been found that DQPTs cannot be identified in the dynamics of the momentum distribution~\cite{Fogarty2017}, whose zero momentum peak can be taken as an order parameter for the equilibrium superfluid phase.

A further class of observables that appear to connect to DQPTs are entanglement quantifiers such as the entanglement entropy~\cite{Jurcevic2016,Schmitt2017} or spin squeezing parameters~\cite{Jurcevic2016}, which is also discussed in more detail in section~\ref{subsec:entanglement}.
Since the Loschmidt amplitude is the Fourier transform of the energy distribution function~\cite{Silva2008gj}, signatures of DQPTs have been also identified in the energy and work statistics~\cite{Heyl2013a,Palmai2015,Abeling2016,Campbell2016}.

\subsection{Classification}
\label{sec:classification}

In equilibrium, phase transitions are grouped into mainly two different categories: First-order phase transitions, associated with a latent heat, and continuous phase transitions that  occur in the presence of a divergent correlation length, not considering for the moment the case of Berezinski-Kosterlitz-Thouless transitions.
Continuous phase transitions are further distinguished in terms of universality classes, entailing all those critical points sharing the same set of critical exponents.
In view of the results summarized in section~\ref{subsec:universality} about scaling and universality, this naturally leads to the question of whether analogous classification schemes might be applicable also to DQPTs.

While there exist DQPTs, which can be classified in the conventional sense because of connections to equilibrium criticality~\cite{Heyl2015dq}, a general classification scheme for DQPTs is not known.
In particular, it is likely that there can be new nonequilibrium fixed points for DQPTs that are not accessible within equilibrium dynamics.
In the infinite-range transverse-field Ising model, for example, it has been found that the DQPT is related neither with the underlying equilibrium quantum nor the thermal phase transitions~\cite{Zunkovic2016,Halimeh2016, Halimeh2017pa}.
In this sense DQPTs represent a critical phenomenon distinct in general from equilibrium phase transitions.
Still, there have been attempts to introduce classification schemes for DQPTs~\cite{Vajna2015,Canovi2014fo}.
For dynamical \emph{topological} quantum phase transitions in noninteracting two-band models, a comprehensive theory of DQPTs has been presented~\cite{Vajna2015}.
In this work it has been shown how DQPTs are related to the underlying equilibrium topological quantum phase transitions.
While so-called topologically-protected DQPTs occur whenever the quantum quench crosses the underlying equilibrium critical point, there can also appear so-called accidental DQPTs which are of genuine nonequilibrium nature, see also section~\ref{subsec:protectedaccidental}.
Furthermore, the work in Ref.~\cite{Vajna2015} has analyzed on general grounds what nonanalytic structures can occur at DQPTs in these models, which can range from simple kink structures to power-laws.
While this work is of central importance for the theory of dynamical topological quantum phase transitions, this classification approach cannot be directly extended to, for example, interacting systems calling for suitable extensions in the future.
An alternative attempt towards a general classification of DQPTs has been put forward in Ref.~\cite{Canovi2014fo}.
Specifically, a definition of first-order DQPTs has been introduced which is based on a new class of observable quantities termed 'generalized expectation values':
\be
	\langle Y(t') \rangle_{\mathcal{G}(t)} = \mathcal{G}^{-1}(t) \langle \psi_0 | e^{-iH(t-t')} Y e^{-iHt'} |\psi_0 \rangle \, ,
\ee
where $\mathcal{G}(t)$ denotes the Loschmidt amplitude and $Y$ a Hermitian operator.
Compared to conventional expectation values these generalized ones only involve a forward time evolution.
Importantly, they can, in principle, be experimentally accessed via an interferometric measurement.
The proposed definition of a first-order DQPT in this work relies on the observation that in the thermodynamic limit $\langle Y(t') \rangle_{\mathcal{G}(t)}$ is dominated by a saddle-point, which, however, doesn't need to stay the same throughout the dynamics.
In this context one can define a first-order DQPT as the point in time where the dominant contribution switches from one to another saddle point.
For those operators $Y$ which are sensitive to the nature of the saddle point, the generalized expectation value $\langle Y(t') \rangle_{\mathcal{G}(t)}$ can exhibit an abrupt change at the critical time, as has then be demonstrated both for a Falicov-Kimball as well as the Hubbard model using dynamical mean-field theory~\cite{Canovi2014fo}.
This definition of a first-order DQPT can be universally applied to any system and does not rely on specific properties of the studied models.
It is, however, not known, how to relate the signatures of first-order DQPTs in these generalized expectation values to conventional measurable quantities such as local observables or correlation functions, which remains an important direction for the future.
In this context it is also of particular interest to study in this light DQPTs in topological systems and therefore to connect to the classification scheme for topological DQPTs discussed before, for which no local observable $Y$ is known to show a signature of DQPTs but rather only global properties.

\section{Further applications and topics}
\label{sec:furtherApplications}

After having discussed the definition of DQPTs and some major characteristic properties, in the following we focus on further applications that have been studied in the recent years.

\subsection{Entanglement dynamics}
\label{subsec:entanglement}

Entanglement has developed into a key concept for the characterization of equilibrium quantum phases and criticality~\cite{Amico2008,LaFlorencie2016}.
It is therefore a natural question whether and how entanglement dynamics and DQPTs are connected to each other.
Such a connection between entanglement dynamics and DQPTs has been observed~\cite{Torlai2014,Canovi2014,Jurcevic2016,Schmitt2017} although the principle underlying this connection is unclear on a general level.
From the phenomenology observed in these works it appears that the concrete relation of DQPTs onto entanglement dynamics can be specific to the nonequilibrium protocol.
For example, in 1D spin chains, evidence has been found that DQPTs can be accompanied by vanishing Schmidt gaps and are thus featured in the entanglement spectrum~\cite{Torlai2014,Canovi2014}, although the connection to DQPTs has not been explored in detail in these works.

\begin{figure}
	\centering
	\includegraphics[width=\columnwidth]{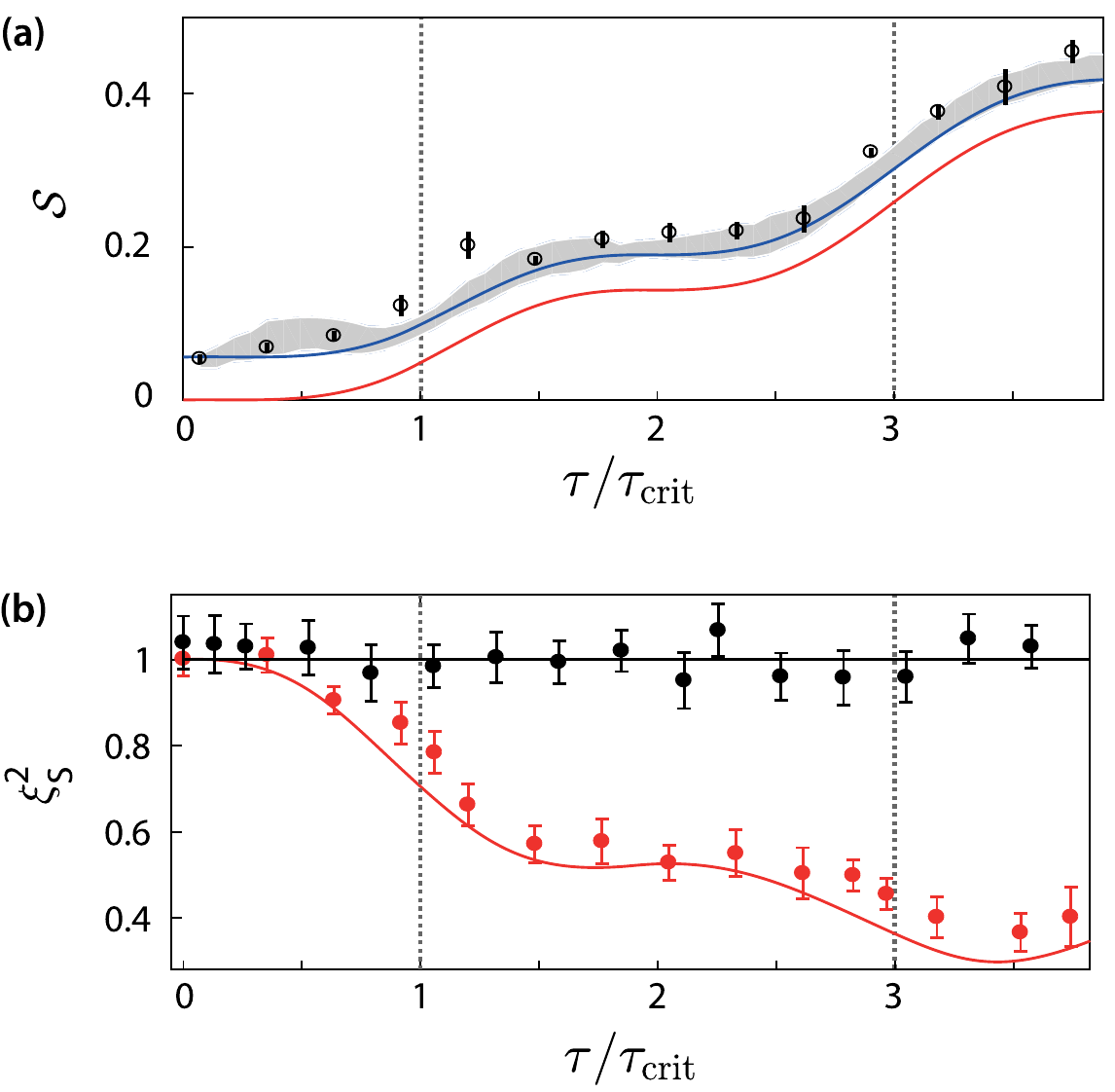}
	\caption{Entanglement dynamics in long-range transverse-field Ising models and the connection to DQPTs as measured in the trapped ion experiment~\cite{Jurcevic2016} for $N=6$ spins and interaction exponent $\alpha\approx 0$. Time is rescaled with respect to $\tau_\mathrm{crit}$ setting the scale of DQPTs appearing in that systems included as dashed lines in the plots. {\bf (a)} The dots represent the measured half-chain entanglement entropy $S$, which is  larger than the expected data (red line) for the ideal evolution. Upon accounting for imperfections in the preparation of the desired fully polarized initial condition (blue line) in combination with uncertainty due to projection noise (shaded grey area) the measured data becomes close to a theoretical estimate. {\bf (b)} The red dots (line) corresponds to the measured (theoretically expected) Kitagawa-Ueda spin squeezing parameter $\xi_s^2$.}
	\label{fig:entanglementDynamicsIons}
\end{figure}
For long-range transverse-field Ising models as realized in the trapped ion experiment, see section~\ref{subsec:expIons} and equation~(\ref{eq:expIonsHamLRIsing}), DQPTs can be associated with a strong entanglement production.
Here, the system is initially prepared in a fully polarized state along the ordering direction of the Ising model and the quantum quench is induced by time-evolving this initial condition with a long-range Ising model at a large value of the transverse field.
In figure~\ref{fig:entanglementDynamicsIons} the measured data of this experiment~\cite{Jurcevic2016} is shown containing both the dynamics of the half-chain entanglement entropy $S$ as well as the Kitagawa-Ueda spin squeezing parameter $\xi_S^2$~\cite{KitagawaUeda}.
As one can see, at the points of DQPTs occurring at $\tau_\mathrm{crit}$ and $3\tau_\mathrm{crit}$ in the units used in the experiment, both the entanglement entropy and spin squeezing exhibit an increased dynamics signaling that the entangling dynamics happens in the vicinity of the DQPTs.
Notice that the lower $\xi_S^2$ the more squeezed and therefore nonclassical the state is.

\begin{figure}
	\centering
	\includegraphics[width=\columnwidth]{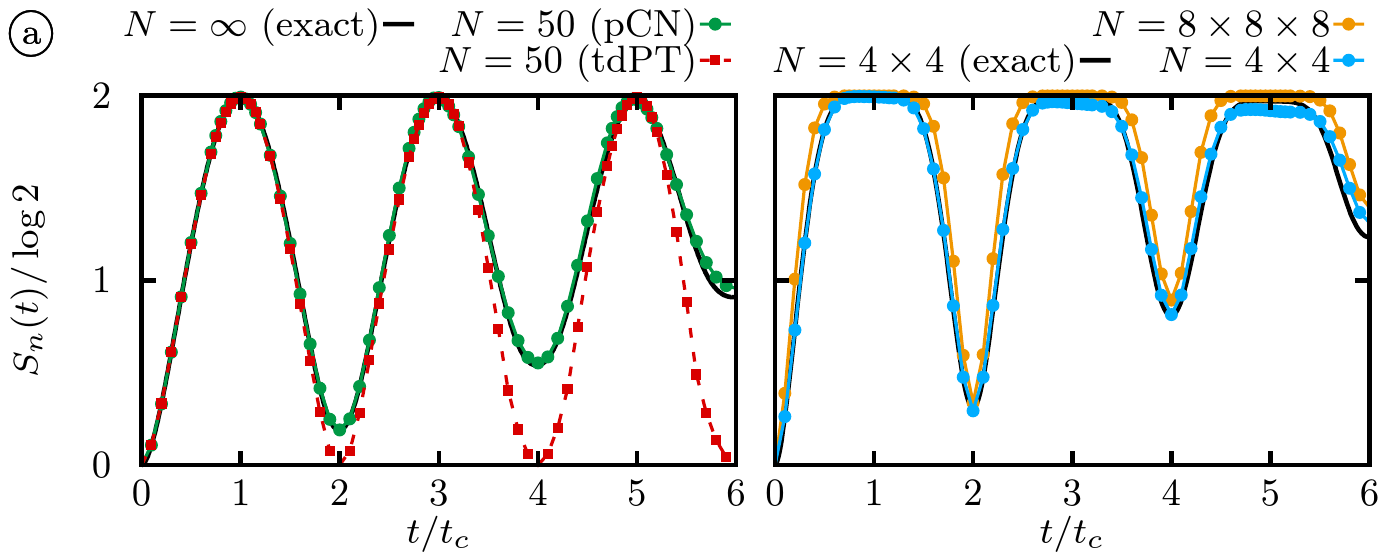}
	\caption{Dynamics of the entanglement entropy $S_n(t)$ for subsystems of size $n=2$ after a quantum quench in short-range transverse-field Ising models~\cite{Schmitt2017}. Here, $t_\mathrm{crit}$ denotes the time scale for DQPTs which appear at $t_c,3t_c,5t_c$. {\bf Left} Comparison of the exact solution with the perturbative classical network (pCN) as well as conventional lowest order time-dependent perturbation theory (tdPT). {\bf Right} Entanglement dynamics for 2D and 3D using the perturbation classical network and comparison to an exact solution for 2D on a $N=4\times4$ lattice.}
	\label{fig:entanglementDynamicsPCN}
\end{figure}
In short-ranged transverse-field Ising models the entanglement dynamics has been recently studied~\cite{Schmitt2017} for a quantum quench already discussed in section~\ref{subsec:universality}.
Contrary to the trapped ion experiment, in this setup the system is initially prepared in a fully polarized state along the transverse-field direction and the quantum quench dynamics is driven by an Ising model for weak transverse field.
In this case the unitary dynamics can be solved for 1D up to 3D by a perturbative mapping to an effective classical network that can be efficiently sampled using Monte-Carlo techniques~\cite{Schmitt2017}.
The results are shown in figure~\ref{fig:entanglementDynamicsPCN}.
As one can see, the entanglement entropy $S_2(t)$ for a subsystem of two nearest-neighboring lattice sites becomes maximal at DQPTs, which are known to occur at times $(2m+1)t_c$ with $m \in \mathbb{N}$ in this model for 1D and 2D.

\subsection{Quantum speed limits}
\label{subsec:quantumSpeedLimit}

The theory of DQPTs has seen a further application in the quantum information context beyond entanglement dynamics, which is quantum speed limits~\cite{Heyl2017a}.
Quantum speed limits provide bounds on the time scale for how fast quantum states can change in real-time evolution~\cite{Mandelstam,Levitin,Campo,Deffner}.
Notice that this need not be speeds associated with the change of local observables or correlations, which are constrained by Lieb-Robinson bounds~\cite{lieb1972,Hastings2006,Nachtergaele2006}.
Quantum speed limits rather quantify the point in time where a state becomes distinguishable from the initial one during progressing dynamics.
Apart from providing limits on the dynamics in closed~\cite{Mandelstam,Levitin} and open~\cite{Campo,Deffner} systems, it has been argued that quantum speed limits might also have applications in optimal control theory~\cite{Caneva} and other quantum technologies such as quantum metrology~\cite{Campo,Deffner}.

Optimal distinguishability is obtained when the time-evolved state becomes orthogonal to the initial one implying a vanishing Loschmidt amplitude $\mathcal{G}(t_c) = 0$ at a time $t_c$.
As discussed extensively in section~\ref{subsec:fisherZeros}, this condition is the defining property of DQPTs providing a general connection between DQPTs and quantum speed limits.
It has been shown that for particular quantum quench protocols, i.e., quenching a quantum critical state by the order parameter, the critical times for DQPTs $t_c$ can exhibit an unconventional system size dependence revealing a yet unrecognized class of quantum speed limits~\cite{Heyl2017a}.

\subsection{Mixed states}
\label{subsec:mixedstates}

The theory of DQPTs has been initially formulated for pure states, see section~\ref{sec:dqpts}.
Recently, extensions to mixed states have been proposed~\cite{Abeling2016,Bhattacharya2017b,HeylBudich2017} which will be discussed in the following.
First of all, it is important to note that the generalization of Loschmidt echos to non-pure states is not unique.
Consequently, each generalization represents a choice targeting to capture certain aspects of the dynamics.
The study in~\cite{Abeling2016} attempted to introduce DQPTs for thermal initial states by interpreting Loschmidt echos as the probability for having performed vanishing work through the quantum quench.
Thus, a natural extension to initial states with nonzero temperature is to compute a work distribution function $P(W,t)$~\cite{Campisi2011fk} and to identify the probability density $P(0,t)$ for having perform no work $W=0$:
\be
	P(0,t)= \sum_{\nu,\mu} \frac{e^{-\beta E_\nu}}{Z}  \left| \langle E_\nu |e^{-iHt} | E_\mu \rangle \right|^2 \delta\big( E_\nu - E_\mu \big) \, ,
\ee
as the generalization with $Z=\sum_\nu e^{-\beta E_\nu}$ the partition function connected to the initial state.
Here, $|E_\nu\rangle$ denotes a complete set of eigenstates of the initial Hamiltonian $H_0$ with associated eigenenergies $E_\nu$.
In the beforementioned article it has then been shown that this definition leads to a smearing of the real-time nonanalyticities in the 1D transverse-field Ising model.
It is, however, unclear how this approach could be formulated for more general initial conditions beyond Gibbs states or nonunitary time evolution because then the notion of a work distribution function faces further challenges~\cite{Campisi2011fk}.
A different route for a theory of DQPTs for mixed states was put forward recently~\cite{Bhattacharya2017b,HeylBudich2017}.
In these works the extension of the Loschmidt amplitude to initial mixed states $\rho_0$ is inspired by an interferometric interpretation of Loschmidt amplitudes which naturally leads to
\be
	\mathcal{G}_\rho(t) = \mathrm{Tr} \big( \rho_0 e^{-iHt} \big) \, ,
\ee
which can be obtained, in principle, from a mixed-state interferometric measurement~\cite{Sjoeqvist2000}.
Alternatively, this extension can be seen as the conventional Loschmidt amplitude for an appropriately purified initial state~\cite{Bhattacharya2017b,HeylBudich2017}.
Using this generalization of Loschmidt amplitudes for quantum quenches in topological systems, DQPTs are not smeared out but can rather persist up to elevated initial temperatures providing a dynamical probe of the topological properties without ever preparing a low-entropy state.
While relaxing the constraint on pureness of the initial conditions, it is important to note that both of the discussed extensions to mixed states require the dynamics to be purely unitary.
This leaves open an important question in the context of experiments because an unavoidable coupling to the environment, though small, induces decoherence and a nonunitary contribution to the system's dynamics becoming relevant at long time scales.

\subsection{General nonequilibrium protocols}
\label{subsec:generalProtocols}

Large parts of this review article are focused on the particular nonequilibrium protocol of a quantum quench.
The natural generalization of Loschmidt amplitudes to other protocols is given by
\be
	\mathcal{G}(t) = \big\langle \psi_0 | U(t) | \psi_0 \big\rangle \, ,
\ee
with $U(t)= \mathcal{T} \exp[-i\int_0^t dt' H(t')]$ denoting the full time evolution operator and $\mathcal{T}$ the time-ordering prescription.
DQPTs in consequence of, e.g., a linear ramp of a parameter instead of a quench have been studied~\cite{Pollmann2010dv,Sharma2016,Puskarov2016,Bhattacharya2017c} where it has been found that DQPTs cannot occur only after the end of nonequilibrium protocol, but also during the parameter ramp.
In this context, one can imagine various further interesting nonequilibrium scenarios such as in the context of periodic Floquet systems or systems subject to noise.

\subsection{Relation to other notions of dynamical phase transitions}
\label{subsec:otherdpts}

In the literature several notions of dynamical phase transitions in quantum many-body systems have been introduced in recent years.
This includes the observation of sudden qualitative changes in the long-time relaxation dynamics in closed quantum systems~\cite{Eckstein2009wj, Garrahan2010xw, Diehl2010, Vosk2014, Smacchia2015,Wang2016,Zhang2017b}, in the non-thermal asymptotic long-time steady states~\cite{Schiro2010gj, Sciolla2010jb, Sciolla2011, Sciolla2013, Maraga2016, Zunkovic2016a}, or also in open quantum many-body systems~\cite{Garrahan2010xw, Diehl2010, Ates2012, Rotter2010, Rotter2015}
Discussing the potential connections of all of these different notions of dynamical phase transitions to DQPTs is beyond the scope of this review.
Therefore, in the following we now summarize those other definitions, for which the relation to DQPTs has already been studied in the literature.
In recent works~\cite{Zunkovic2016,Wang2017,Weidinger2017} connections between DQPTs and such other notions of nonequilibrium criticality have been addressed.
The nonequilibrium phase transition discussed in reference~\cite{Zunkovic2016} concerns a long-range transverse-field Ising model along the lines of the trapped ion experiment discussed in section~\ref{subsec:expIons}.
Preparing the system in a fully polarized state along the ordering direction and performing the dynamics, this system supports a non-thermal steady state transition.
When the transverse field is weak, the system ends up in a phase with a nonzero magnetization, whereas for large fields the magnetization vanishes~\cite{Sciolla2011}. 
These two phases are separated by a nonequilibrium phase transition that has no equilibrium counterpart~\cite{Sciolla2011}.
This type of dynamical phase transition appears in a model system which is known to also exhibit a so-called excited-state quantum phase transition~\cite{Cejnar2007, Caprio2008,ExcState1,Santos2015,Santos2016, Stransky2016, Sindelka2017}.
To which extent these are related remains at the current point an interesting and open question.
As it has been shown in~\cite{Zunkovic2016}, this transition in the steady state is connected to a DQPT in the full ground state return probability, see equation~\ref{eq:groundStateProb}, occurring on transient to intermediate time scales.
How the anomalous DQPTs~\cite{Halimeh2016,Halimeh2017pa,Homrighausen2017}, that occur for weak transverse fields at long times, can be merged into this picture is, however, not yet known.
A generalization to the case of a system, exhibiting a broken continuous symmetry, has been recently studied for the nonequilibrium dynamics in an $O(N)$-model~\cite{Weidinger2017}.
In contrast to the case of a broken $\mathbb{Z}_2$ symmetry in the Ising model, the real-time nonanalyticity of the DQPTs doesn't appear in the leading order in $n$ (number of degrees of freedom) contribution to the rate function $\lambda(t)$, but rather in a $n^{-1}$ correction similar to what happens at an equilibrium surface phase transitions.

A connection between DQPTs and a nonequilibrium topological transition in the steady state after a quench has been identified recently in reference~\cite{Wang2017}.
As opposed to the Ising model whose steady-state transition is characterized by a local order parameter, in this topological case it is signaled by a nonanalytic behavior in linear-response transport properties~\cite{Wang2016}.
What has been observed is that DQPTs appear if an only if the control parameters are quenched across the critical value of the steady state transition.
To which extent DQPTs can also be connected to other notions of nonequilibrium criticality remains an open question.

\subsection{Inhomogeneous systems}
\label{subsec:inhomogeneous}

Dynamical quantum phase transitions have been studied also for inhomogeneous systems including the random energy model~\cite{Takahashi2012, Takahashi2013}, the Anderson model~\cite{Yin2017}, as well as for systems in incommensurate lattices in the context of the Aubry-Andre model~\cite{Yang2017}.
While for the random energy model and the Aubry-Andre model strong evidence for DQPTs has been observed, in the Anderson model only sharp but still smooth structures have been found for large but finite systems leaving open the question of how these sharp features behave in the thermodynamic limit.
It is currently unclear to which extent disorder influences DQPTs on a general level.
This is of particular importance because disorder can also have a drastic influence for equilibrium phase transitions.
In particular, a large class of first-order phase transitions is smeared in the presence of disorder and develops into a smooth crossover~\cite{Imry1975, Aharony1976, Aizenman1989, Aizenman1990}.
For continuous transitions, the Harris criterion can predict the stability of a transition against weak disorder on the basis of the critical exponents of the transition~\cite{Harris1974}.
To which extent these theorems can be extended to DQPTs is not yet known.

\section{Prospects}
\label{sec:propects}

As summarized in this review, the field of DQPTs has seen substantial progress in recent years both from the theoretical as well as experimental point of view.
While this progress underlines the potential of the theory of DQPTs to provide a principle for the understanding of the dynamics in quantum many-body systems, some major questions are still open.
It is the aim of this final section to devise some of the challenges in this context, to turn to potential prospects, and to point out directions of future research on DQPTs.

One of the main challenges within the theory of DQPTs is the lack of the notion of a free energy for the considered nonequilibrium quantum states.
As already anticipated in the introduction, this lack, however, might not only be seen as an obstacle but also as the defining property of the considered nonequilibrium quantum states providing also the room for properties inaccessible within equilibrium thermodynamics.
Still, major questions remain:
Is there nevertheless a macroscopic description?
Is it possible to construct a nonequilibrium counterpart of a Landau-Ginzburg theory?
Associated with that:
Is there an organizing principle analogous to the minimization of free energies?
Addressing these questions clearly constitutes one of the most pertinent tasks in the theory of DQPTs.
Up to now, the majority of the works on DQPTs have seen exponents which are integer valued.
Although this should not be interpreted such that the nature of the considered DQPTs is of first order in the equilibrium sense, see e.g. reference~\cite{Heyl2015dq}, it is not clear whether more exotic DQPTs can exist displaying nontrivial exponents.
This might call for novel approaches that allow to compute Loschmidt amplitudes for higher-dimensional interacting theories.
Fortunately, DQPTs occur on transient time scales where approximative methods can be much better controlled as opposed to the case when studying the long-time dynamics of correlated systems.
A further potential scope of the theory of DQPTs is to explore connections to other nonequilibrium phases and critical phenomena including fields such as many-body localization (MBL)~\cite{Schreiber2015oo,Smith2016,Bordia2016,Choi2016,Altman2014,Nandkishore2015}, or nonthermal fixed points~\cite{Berges2008,Berges2009}.
In this context, it might be of particular interest to study through the lense of DQPTs so-called eigenstate phases~\cite{Huse2013,Moessner2017}.
Remarkably, these eigenstate phases cannot only be associated with long-range spatial correlations such as in the case of MBL spin glasses~\cite{Huse2013,Vosk2014,Kjaell2014}, but also with unconventional spatio-\emph{temporal} order for time crystals~\cite{Wilczek2012,Khemani2016,Else2016,Choi2017,Zhang2017}.
From a more general point of view, the theory of DQPTs captures nonanalyticities in the time translation operator.
In this light a natural question is to which extent analogs to these nonanalyticities are possible in different kinds of translations or rotations as they appear, for example, in the many-particle momentum-translation operator~\cite{KingSmith1993,Resta1998,Bardyn2017} which plays a central role for topological systems.

\ack
The author would like to thank for stimulating discussions and collaborations on topics related to this review with various colleagues including Rainer Blatt, Jan Budich, Sebastian Diehl, Nick Fl\"aschner, Philipp Hauke, Yi-Ping Huang, Petar Jurcevic, Ben Lanyon, Achilleas Lazarides, Stefan Kehrein, Michael Knap, Marcus Kollar, Roderich Moessner, Anatoli Polkovnikov, Christian Roos, Sthitadhi Roy, Markus Schmitt, Alessandro Silva, Daniele Trapin, Dominik Vogel, Matthias Vojta, Simon Weidinger, Christof Weitenberg, Peter Zoller, and Bojan Zunkovic. Financial support by the Deutsche Forschungsgemeinschaft via the Gottfried Wilhelm Leibniz Prize program is gratefully acknowledged.

\section*{References} 

\bibliographystyle{iopart-num}
\bibliography{literature}

\end{document}